\documentclass[12pt]{article}
\usepackage[utf8]{inputenc}

\usepackage[a4paper,top=3cm,bottom=3cm,left=3cm,right=3cm]{geometry}

\usepackage[normalem]{ulem}

\usepackage{amssymb}
\usepackage{amsmath}
\usepackage{amsfonts}
\usepackage{booktabs}
\usepackage{xcolor}
\usepackage{float}
\usepackage{subfig}
\usepackage{graphics}
\usepackage{graphicx}
\usepackage{enumitem}

\usepackage{amsthm}
\usepackage{listings}
\theoremstyle{definition}

\usepackage{natbib}

\usepackage{hyperref}
\hypersetup{
citecolor= blue, 
    colorlinks=true,
    linkcolor=blue, 
    filecolor=blue,      
    urlcolor=blue,
    pdftitle={Overleaf Example},
    pdfpagemode=FullScreen,
    }
\usepackage{framed,color,verbatim}
\definecolor{shadecolor}{rgb}{.95, .95, .95}
   {\snugshade\verbatim}%
   {\endverbatim\endsnugshade}

\DeclareMathOperator{\e}{e}
\newcommand{\de}{\mathrm{d}}

\date{}

\begin{document}

\title{
Semi-parametric profile pseudolikelihood
via local summary statistics for 
spatial point pattern intensity estimation}

 \author{Nicoletta D'Angelo$^{*,1}$, Giada Adelfio$^{1}$, Jorge Mateu$^{2}$, Ottmar Cronie$^{3}$}

\date{1. Department of Business, Economics and Statistics, University of Palermo, Palermo, Italy;\\ 2. Department of Mathematics, University Jaume I, Castellon, Spain;\\ 3. Department of Mathematical Sciences, Chalmers University of Technology and University of Gothenburg, Gothenburg, Sweden.}

\maketitle

\begin{abstract}
Second-order statistics play a crucial role in analysing point processes. Previous research has specifically explored locally weighted second-order statistics for point processes, offering diagnostic tests in various spatial domains. However, there remains a need to improve inference for complex intensity functions, especially when the point process likelihood is intractable and in the presence of interactions among points. This paper addresses this gap by proposing a method that exploits local second-order characteristics to account for local dependencies in the fitting procedure. Our approach utilises the Papangelou conditional intensity function for general Gibbs processes, avoiding explicit assumptions about the degree of interaction and homogeneity. We provide simulation results and an application to real data to assess the proposed method's goodness-of-fit. Overall, this work contributes to advancing statistical techniques for point process analysis in the presence of spatial interactions.

\medskip
{\bf Keywords}: Intensity estimation, Local characteristics, Pseudolikelihood, Spatial point patterns, Summary statistics

\end{abstract}

\section{Introduction}

When dealing with point process statistics, a main interest is often to infer how available covariates influence the expected abundance of events, expressed as a function of space. More specifically, one wants to model the intensity function of the underlying point process in terms of the available spatial covariates. 
Since likelihood functions for point processes are typically intractable \citep{van2000markov}, the typical way to handle this is to use composite likelihood estimation, which simply amounts to maximising a Poisson process likelihood function, regardless of whether the underlying point process is a Poisson process or not \citep{guan2006composite,lindsay1988composite,bayisa2023regularised}. 
Although this has nice asymptotic properties, in terms of consistency, for sufficiently nice point processes \citep{schoenberg2005consistent}, it amounts to the inherent dependence structure being asymptotically negligible. 
However, since the asymptotic results do not incorporate rates of convergence, they are of limited practical use. 
In addition, for normal-sized point patterns, which is what one is commonly dealing with 
in real-life applications, 
the underlying dependence structure certainly has negative effects on the outcome, when using a Poisson process likelihood function to model the intensity of a non-Poissonian point process. 
Composite likelihood estimation, disregarding the dependence structure in the underlying point process, increases the risk of over-fitting \citep{cronie2018non}, which is manifested by 
the statistical procedure interpreting all aggregations of points in a point pattern as the effect of local intensity function peaks, i.e.\ influence of covariates, instead of the combination of both influence of covariates and spatial dependence (clustering/attractiveness). 
Put differently, we observe the effects of both spatially structured covariates and random effects, and a Poisson process-based procedure cannot appropriately distinguish between the two based on only one realisation of the point process. 
Hence, ideally we would have that the statistical procedure employed for the intensity estimation adjusts for the inherent spatial dependence in some suitable way. Depending on how this is done, the procedure is either fully parametric or semi-parametric. In the fully parametric setting an alternative could be quasi-likelihood estimation \citep{guan2015quasi}, which has composite likelihood estimation as a special case.

In this work we opt for a semi-parametric approach, and we account for the dependence structure by applying appropriate weighting using an estimated second-order interaction term. More specifically, we start from the formulation of a Gibbs process, whose (Papangelou) conditional intensity may be expressed as the product of two terms: the first corresponds to a Poisson process intensity function, which incorporates the covariates (in a log-linear way), and the second term adjusts the first term with respect to the actual dependence structure of the Gibbs process. Our idea is effectively to estimate the second term, the interaction term, externally and then plug it into the conditional intensity function. Exploiting that the intensity function is given by the expected value of the conditional intensity function, we here propose a way to estimate the intensity function by means of weighted composite likelihood estimation, where we weight the parametric intensity form by a non-parametric estimate of the expectation of the interaction term.

The non-parametric estimate of the expectation of the interaction term can be obtained in a number of ways and here we propose basing it on an estimate of the interaction structure, such as the pair correlation function or the $K$-function. Although the literature on summary statistics is vast \citep{baddeley2000non,iftimi2019second,d2023local,cronie:lieshout:15},
as local summary statistics provide further insight into the local structure of the underlying point process, we argue that the natural way forward is exploiting second-order local summary statistics, so called local indicators of spatial association (LISA functions). This is in line with our previous works \citep{adelfio2020some,dangelo2021local}, where a weighted version of second-order statistics for spatio-temporal point processes was used to provide diagnostic tests in different domains. 

Although our methodology is presented for point processes/patterns in Euclidean domains, it should be noted that it may be straightforwardly applied to non-Euclidean settings as well, e.g.\ linear networks where we may make use of e.g.\ the local higher-order summary statistics of \citet{cronie2020inhomogeneous}.

The structure of the paper is as follows. Section \ref{sec:preliminaries} contains some brief preliminaries on spatial point processes and introduces the Papangelou conditional intensity, while setting the notation.  
Section \ref{sec:proposal} is devoted to our methodological proposal, considering the local summary statistics in the penalisation terms of the composite likelihood. Simulation results for assessing the goodness-of-fit of the proposed method are reported in Section \ref{sec:sims}. An application is presented in Section \ref{sec:appl} to demonstrate the applicability of the method further. Conclusions are drawn in Section \ref{sec:conc}.

\section{Point process preliminaries}\label{sec:preliminaries}

We consider a (simple) point process $X=\{x_i\}_{i=1}^N$ 
\citep{daley:vere-jones:08}, with points $x_i$ in a subset $W$ of the two-dimensional Euclidean space $\mathbb{R}^2$, which is equipped with 
the Lebesgue measure $\vert A \vert =\int_A\text{d}z$ for Borel sets $A\in\mathcal{B}(\mathbb{R}^2)$; a closed Euclidean $r$-ball around $x\in\mathbb{R}^2$ will be denoted by $b[x,r]=\{y\in\mathbb{R}^2:\|x-y\|\leq r\}$. Formally, $X$ is a random element in the measurable space $\mathcal{N}$ of locally finite point configurations/patterns $\mathbf{x}=\{(x_1,\ldots, x_n)\}$, $n\geq0$ \citep{van2000markov}. A point location in the two-dimensional plane is denoted by a lowercase letter such as $u$, and 
it can be specified by its Cartesian coordinates, i.e.\ $u = (u_1 ,u_2 )$, in such a way that we do not need to mention the coordinates explicitly.

\subsection{Papangelou conditional intensity
}

The Papangelou conditional intensity of a point process $X$ 
satisfies 
\citep{van2000markov}
\[
\lambda_{\theta}(u;\mathbf{x})
=
\mathbb{P}(X\cap du\neq\emptyset|X\cap du^c = \mathbf{x}\cap du^c)/\de u,
\qquad \theta\in\Theta\subseteq\mathbb{R}^{d'}, u\in W, 
\]
where $\mathbf{x}\in\mathcal{N}$ is an arbitrary point pattern contained in the observation window $W$, provided that $X$ is generated by the parametrisation $\theta$. In words, 
$\lambda_{\theta}(u;\mathbf{x})$ is the density of the conditional probability of finding a point of $X$ in an infinitesimal neighbourhood $du$ of $u$, given that $X$ coincides with $\mathbf{x}$ outside $du$. 

In the case of Gibbs models, which include e.g.\ Markov point processes, Cox process and Hawkes processes, the Papangelou conditional intensity is conveniently expressed as 
\begin{equation}
   \lambda_{\theta}(u;\mathbf{x}) = \e^{\phi_{\theta}^1(u) + \phi_{\theta}^2(u,\mathbf{x})}
={\nu}_{\theta}(u)\e^{\phi_{\theta}^2(u,\mathbf{x})},
\label{eq:PapangelouCIF}
\end{equation}
where $\e^{\phi_{\theta}^2(u,\mathbf{x})}$ may be viewed as a dependence adjustment with respect to a Poisson process with intensity function ${\nu}_{\theta}(u) = \e^{\phi_{\theta}^1(u)}$; this means that $\phi_{\theta}^2(\cdot)=0$ for a Poisson process. 
Moreover, a model is called attractive (or repulsive) if when $\mathbf{x}\subseteq\mathbf{y}$ we have that $\lambda_{\theta}(u;\mathbf{x})\leq \lambda_{\theta}(u;\mathbf{y})$ (or $\lambda_{\theta}(u;\mathbf{x})\geq \lambda_{\theta}(u;\mathbf{y})$); following \eqref{eq:PapangelouCIF}, this is equivalent to $\phi_{\theta}^2(u,\mathbf{x})\leq \phi_{\theta}^2(u,\mathbf{y})$ (or $\phi_{\theta}^2(u,\mathbf{x})\geq \phi_{\theta}^2(u,\mathbf{y})$). 
This means that a Poisson process is both attractive and repulsive. 
An explicit example here is a pairwise interaction process, where $\phi_{\theta}^2(u,\mathbf{x}) = \sum_{x\in\mathbf{x}}\bar\phi_{\theta}^2(u,x)$ for a symmetric function $\bar\phi_{\theta}$, which is typically only distance dependent, i.e.\ $\bar\phi_{\theta}(u,x)=\bar\phi_{\theta}^2(\|u-x\|)$ \citep[Section 6.2]{moller:waagepetersen:04}. 
Given a set of covariates, $z(u)=(z_1(u),\ldots,z_d(u))$, $u\in W$, over the observation window $W$, in modelling settings it is common to assume that $\phi_{\theta}^1(u)=\theta_0 + \theta_1 z_1(u) + \cdots + \theta_d z_d(u)$, where $(\theta_0,\ldots,\theta_d)\in \mathbb{R}^{d+1}$ is a part of the complete parameter vector $\theta\in\Theta$, and then
\begin{equation}
   {\nu}_{\theta}(u)
=\exp\{\theta_0 + \theta_1 z_1(u) + \cdots + \theta_d z_d(u)\}.
\label{eq:nu}
\end{equation}
We further have that the (first-order) intensity function corresponding to \eqref{eq:PapangelouCIF} is given by
\begin{equation}
\rho_{\theta}(u)
=
E[\lambda_{\theta}(u;X)] 
= 
{\nu}_{\theta}(u)
E[\e^{\phi_{\theta}^2(u,X)}]
=
{\nu}_{\theta}(u)
\phi_{\theta}^*(u)
.
    \label{eq:PapangelouInt}
\end{equation}

\subsection{Intensity estimation}

For many models, even if $\phi_{\theta}^2$ is known, the second term on the right-hand side of \eqref{eq:PapangelouInt}, $\phi_{\theta}^*(u)$, is unknown/intractable, whereby exact intensity estimation is infeasible. The solution proposed in the literature is often to set $\phi_{\theta}^*(u)=1$ and proceed as if one is dealing with a Poisson process, i.e.\ to find an estimate by maximising the Poisson log-likelihood
\begin{equation}
    \theta\mapsto
\sum_{x\in\mathbf{x}}
\log\rho_{\theta}(x)
-
\int_W \rho_{\theta}(u) \de u
=
\sum_{x\in\mathbf{x}}
\log{\nu}_{\theta}(x)
-
\int_W {\nu}_{\theta}(u) \de u.
\label{eq:CLE}
\end{equation}
This approach is referred to as composite likelihood estimation. 
However, this assumption becomes problematic if $\phi_{\theta}^*(u)$ deviates significantly from 1, which implies that there are stronger interactions at large. By not adjusting the intensity for the intrinsic dependence structure, one runs the risk of overfitting. This can lead to an over-belief in the influence of covariates on the outcome of the point process. Here, we naturally have the option of replacing the parametrised intensity function $\rho_{\theta}$ in \eqref{eq:CLE} by the parametrised Papangelou conditional intensity $\lambda_{\theta}$, 
which results in pseudolikelihood estimation \citep{van2000markov}. However, this requires the selection of a specific family of parametric models, a decision that may not be easy or even necessary, especially when our goal is to only obtain a parametric estimate of the intensity function, to understand the effects of covariates on the underlying point process. Note that the choice of a parametric model family to use for $\lambda_{\theta}$ commonly involves non-parametric analyses of inhomogeneous summary statistics. These analyses help to identify whether the underlying point process shows clustering, inhibition, or a completely random structure. However, these approaches have limitations, as they often do not provide specific insights into the exact model family that would be most appropriate. 

\section{Locally penalised Poisson log-likelihood}\label{sec:proposal}

\citet{cronie2023cross} noted that the approach to non-parametric intensity estimation introduced by 
\citet{cronie2018non} actually entailed treating a kernel intensity estimator as a conditional intensity function. More specifically, they noted that since any non-parametric intensity estimator $\widehat\rho_{\theta}(u,\mathbf{x})$, $u\in W$, $\mathbf{x}\in\mathcal{N}$, with tuning parameter $\theta$ is a non-negative measurable function on $W\times \mathcal{N}$, it formally satisfies the properties of being a conditional intensity model. Based on this, \citet{cronie2023cross} argued as follows. Given a method to fit a parametrised conditional intensity to a point pattern $\mathbf{x}$, we can obtain an estimate $\widehat\theta=\widehat\theta(\mathbf{x})$ by minimising the associated loss function $\mathcal{L}(\theta)=\mathcal{L}(\widehat\rho_{\theta},\mathbf{x})$. Provided that this method does a good job, we have that $\widehat\rho_{\widehat\theta}$ should be close to the true data-generating conditional intensity model $\lambda$. Since we apply the same conditional intensity model $\widehat\rho_{\theta}$ regardless of whether there in fact is a $\theta$ such that $\widehat\rho_{\theta}=\lambda$, we are typically in the setting of a misspecified model. Now, if the realisation $\mathbf{x}$ is central in the distribution corresponding to $\lambda$, we have $\rho(\cdot) = E[\lambda(\cdot;X)]\approx \lambda(\cdot;\mathbf{x}) \approx  \widehat\rho_{\widehat\theta}(\cdot;\mathbf{x})$ for the true intensity function $\rho(\cdot)$. Hence, we fit a non-parametric intensity estimator by treating it as a conditional intensity and by plugging the observed point pattern into the resulting estimate we obtain an estimate of the intensity function. 

Following this reasoning, our idea here is to: 
\begin{enumerate}
    \item Specify an estimator $\widehat\phi^2$ for the interaction function in \eqref{eq:PapangelouCIF} and treat $\widehat\phi^*(u)=\e^{\widehat\phi^2(u,\mathbf{x})}$, where $\mathbf{x}$ is the observed point pattern, as an estimate of the expectation in \eqref{eq:PapangelouInt}.
    \item Plug this into the intensity expression \eqref{eq:PapangelouInt}, whereby we are left with fitting ${\nu}_{\theta}(x)\widehat\phi^*(x)$ to $\mathbf{x}$, 
    where ${\nu}_{\theta}$ is given by \eqref{eq:nu}. In order to do so, we turn to pseudolikelihood estimation, which as a result amounts to maximising the weighted Poisson log-likelihood  
    \begin{equation}
    \theta\mapsto
    \sum_{x\in\mathbf{x}}
    \log({\nu}_{\theta}(x)
    \widehat\phi^*(x))
    -
    \int_W {\nu}_{\theta}(u)
    \widehat\phi^*(u) du.
    \label{eq:CLEpen}
    \end{equation}
    Rewriting \eqref{eq:CLEpen} as
\begin{equation}
 \theta\mapsto\sum_{x\in\mathbf{x}}
\log{\nu}_{\theta}(x)
-
\int_W {\nu}_{\theta}(u) \de u
+
\left[
\sum_{x\in\mathbf{x}}
\log\widehat\phi^*(x)
+
\int_W {\nu}_{\theta}(u)(1-\widehat\phi^*(u)) \de u
\right]
,
 \label{eq:CLEpen2}
\end{equation}
we see that it may be viewed as a regularised form of a Poisson log-likelihood, which is penalised by the amount of interaction we have. 
\end{enumerate}

Ideally, we expect a clustered/attractive process to have $\widehat\phi^*(u)>1$ on average, while for a Poisson process we aim to have $\widehat\phi^*(u)=1$, and $\widehat\phi^*(u)<1$ for an inhibitory/repulsive process. Since many pairwise interaction-based Gibbs/Markov point processes satisfy that $\phi_{\theta}^2(u,\mathbf{x})$ counts the number of pairs of points which satisfy a certain closeness criterion \citep{van2000markov}, it is quite natural to design an estimator that reflects this behaviour. Moreover, we also aim to capture local variations in the dependence structure, i.e.\ a structure varying in space. Therefore, we find it natural to make use of local summary statistics. More specifically, for any $x_i\in\mathbf{x}$ we let
\begin{equation}
  \widehat\phi^*(x_i)
=
\widehat\phi^*(x_i;\mathbf{x}\setminus\{x_i\})
=
t(\hat{K}^i, K_{Pois}),
\label{eq:phistar}
\end{equation}
with $t(\cdot,\cdot)$ a {\em discrepancy measure} quantifying the discrepancy between the local estimator $\hat{K}^i=\{\hat{K}^i(r)\}_{r\geq0}$ and its theoretical counterpart $K_{Pois}=\{K_{Pois}(r)\}_{r\geq0}$. 
Here, $\hat{K}^i(r)$ is the local $K$-function estimate associated to $x_i\in\mathbf{x}$, given by 
\begin{equation}
\label{e:GeneralLocal}
 \hat{K}^i(r)
  = \frac{1}{\vert W \vert}
  \sum_{x_j\in \mathbf{x} 
  }^{\ne}
  w(x_i,x_j)\mathbf{1}\{x_j\in x_i+C\}/(\tilde{\rho}(x_i)\tilde{\rho}(x_j))
  ,
\end{equation}
 where $\neq$ indicates that the sum is over couples of distinct points of  $\mathbf{x}$, $W\subseteq\mathbb{R}^2$, $\vert  W \vert  >0$, and $w(\cdot)$ is an edge correction term \citep{d2023local}. 
When $\tilde{\rho}(\cdot)$ is set to the true intensity $\rho>0$ of a homogeneous point process $X$, and $\mathbf{x}$ is set to $X\cap W$, \eqref{e:GeneralLocal} reduces to a local estimator of Ripley's $K$-function, provided that $x_i+C=x+b[0,r]=b[x_i,r]$. Conversely, if the process is inhomogeneous, if we let $\tilde{\rho}(\cdot)$ be an estimate of the underlying intensity function, 
\eqref{e:GeneralLocal} corresponds to a local estimator of the inhomogeneous $K$-function \citep{baddeley2000non}.  Here, we employ the stationary version of the $K$-function because at small ranges $r$ the homogeneous and the inhomogeneous versions of the $K$-function are not expected to deviate much. 
It is worth pointing out that we here are effectively dealing with a functional marked point process \citep{ghorbani2021functional, d2023local}, which is again transformed into a marked point process with real-valued marks, using the discrepancy functional to achieve the presented transformation.
 
The theoretical value of \eqref{e:GeneralLocal} for a Poisson process is $K_{Pois}(r) = \pi r^2$ when $\tilde{\rho}(\cdot)=\rho$ and $W\subseteq\mathbb{R}^2$ \citep{getis1984interaction}. 
Hence, if $\hat{K}^i$ wiggles around $K_{Pois}$, we have that $\widehat\phi^*(x_i)\approx1$, in such a way that equation \eqref{eq:CLEpen2} (approximately) reduces to the Poisson log-likelihood \eqref{eq:CLE}, basically giving a penalty equal to zero. On the other hand, if $\hat{K}^i$ is systematically larger (or smaller) than $K_{Pois}$, whereby we have local clustering (or inhibition), then we have $\widehat\phi^*(x_i)>1$ (or $\widehat\phi^*(x_i)<1$). 

\subsection{Interaction function interpolation}

Regardless of the choice of $t(\cdot,\cdot)$, which will be discussed in Section \ref{s:Discrepancies}, in order to obtain an estimate $\widehat\phi^*(u)$ for all locations $u\in W$, we need to use the collection $\{\widehat\phi^*(x_i):x_i\in\mathbf{x}\}$ to generate interpolations $\{\widehat\phi^*(u):u\in W\setminus\mathbf{x}\}$. Note that we need the full surface $\{\widehat\phi^*(u):u\in W\}$ to compute the integral in equation \eqref{eq:CLEpen}. In practical terms, the optimisation of \eqref{eq:CLEpen} is achieved by first considering
\begin{equation}
   {\widetilde \nu}_{\theta}(u)
   =
   \exp\{B(u)\}{\nu}_{\theta}(u)
   =
    \exp\{\theta_0 + \theta_1 z_1(u) + \cdots + \theta_d z_d(u) + B(u)\},
\label{eq:nu_offset}
\end{equation}
i.e.\ introducing an external offset $B(u)=\widehat\phi^*(u)$, $u\in W$, into \eqref{eq:nu}, and then optimising the Poisson log-likelihood \eqref{eq:CLE} with ${\nu}_{\theta}$ replaced by ${\widetilde \nu}_{\theta}$. This is extremely convenient, as it allows us to make use of efficient standard software for Poisson process intensity modelling, most notably the function \texttt{ppm.ppp} in the \texttt{R} package \texttt{spatstat} \citep{baddeley:rubak:tuner:15}.

There are different approaches to carry out the spatial interpolation. Here we will focus on spatial smoothing of the numeric values observed at the point pattern locations, where
$$
B(u) = \frac{\sum_{i=1}^n w_i(u) \hat{\phi^*}(x_i)}{\sum_{i=1}^n w_i(u)},
$$
given the estimates \(\hat{\phi^*}(x_1),\ldots,\hat{\phi^*}(x_n)\) at the point pattern locations \(\mathbf{x}=\{x_1,\ldots,x_n\}\), together with a collection of suitable weight functions $w_i$, $i=1,\ldots,n$. 

The specific case of kernel weighting, using the Gaussian kernel $\kappa$, is achieved by letting $w_i(u)=\kappa(u-x_i)$. This yields the Nadaraya-Watson smoother \citep{nadaraya1964estimating,nadaraya1989nonparametric,watson1964smooth}
$$
B_{kernel}(u) = \frac{\sum_{i=1}^n \kappa(u-x_i) \hat{\phi^*}(x_i)}{\sum_{i=1}^n \kappa(u-x_i)}
$$
for any location $u\in W\setminus\mathbf{x}$. Here, the smoothing kernel bandwidth is chosen by least squares cross-validation.

An alternative is inverse-distance weighting \citep{shepard1968two}, where 
each weight is given by 
$w_i(u)=\|u-x_i\|^{-p}$, i.e.\ the inverse $p$th power of the Euclidean distance between $u$ and $x_i$. Consequently, the smoothed value at $u$ is given by
$$
B_{idw}(u) = \frac{\sum_{i=1}^n \|u-x_i\|^{-p} \hat{\phi^*}(x_i)}{\sum_{i=1}^n \|u-x_i\|^{-p}}.
$$

\subsection{Specific discrepancies}
\label{s:Discrepancies}

Formally, the discrepancy is a functional, taking two functions as inputs and providing a real number. Clearly, we want $t(f,f)=0$ for any function $f$. A natural starting point would be a metric reflecting distances between functions. Two such options would be $t(\hat{K}^i, K_{Pois}) = \sup_{r\in[r_0,r_{max}]}|\hat{K}^i(r)-K_{Pois}(r)|$ and $t(\hat{K}^i, K_{Pois}) = (\int_{r_0}^{r_{max}}(\hat{K}^i(r)-K_{Pois}(r))^2\text{d}r)^{1/2}$, i.e.\ the uniform metric and the $L_2$ metric for functions $f:[r_0,r_{max}]\to\mathbb{R}$, given suitable spatial range value bounds $r_0$ and $r_{max}$. 
Another intuitive proposal for $t(\cdot,\cdot)$ is the squared difference
\begin{equation}
t(\hat{K}^i, K_{Pois})
=
\exp \Bigg\{\int_{r_0}^{r_{max}}\Big(\hat{K}^i(r)-K_{Pois}(r)\Big)^2\text{d}r\Bigg\}.
\label{eq:phistar2}
\end{equation}
It turns out, however, that these choices put too little weight on the spatial dependencies for \eqref{eq:CLEpen} to deviate enough from \eqref{eq:CLE} when the underlying point process, in fact, exhibits spatial interaction.

\begin{figure}[H]
     	\includegraphics[width=.5\textwidth]{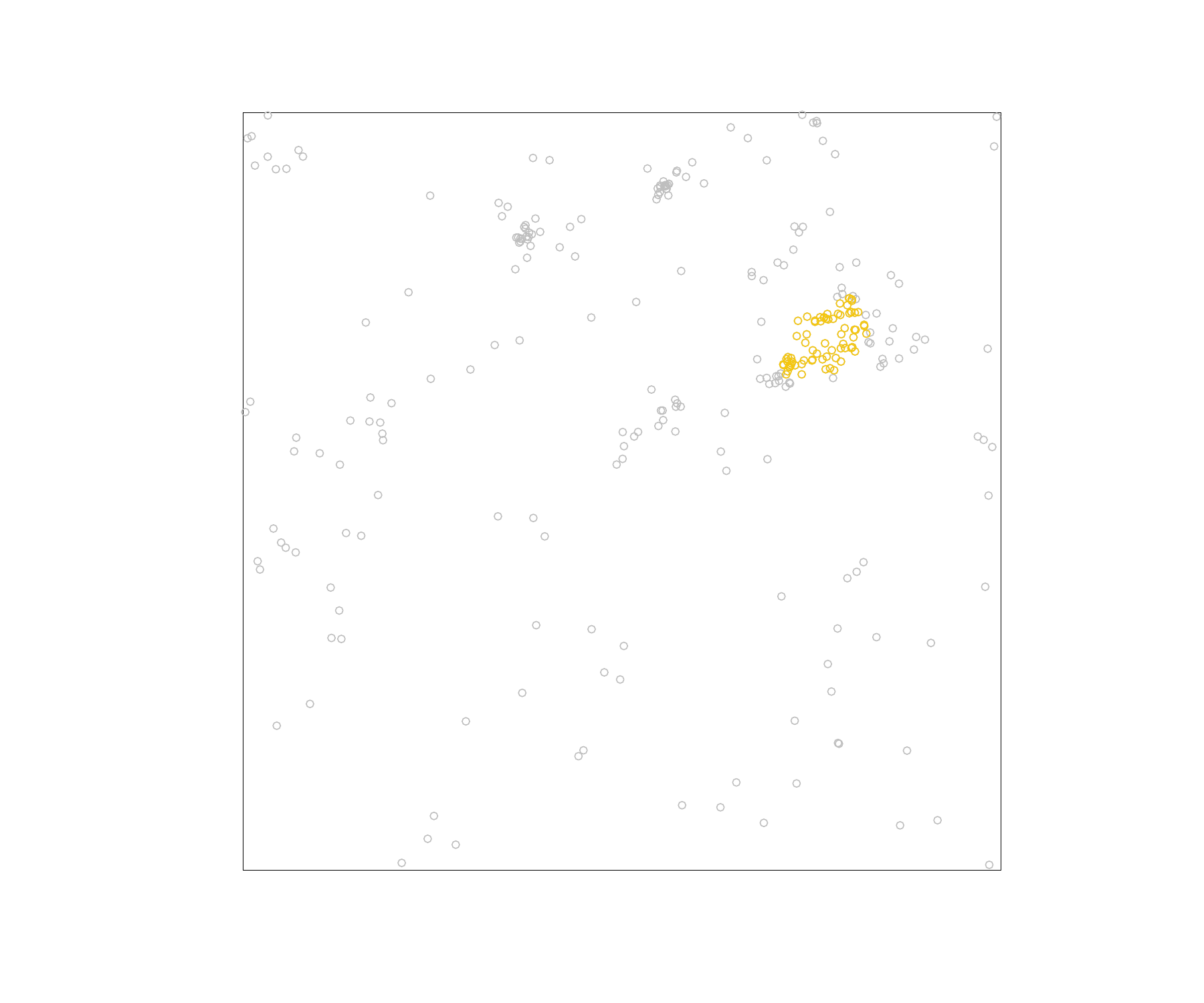}
 \includegraphics[width=.5\textwidth]{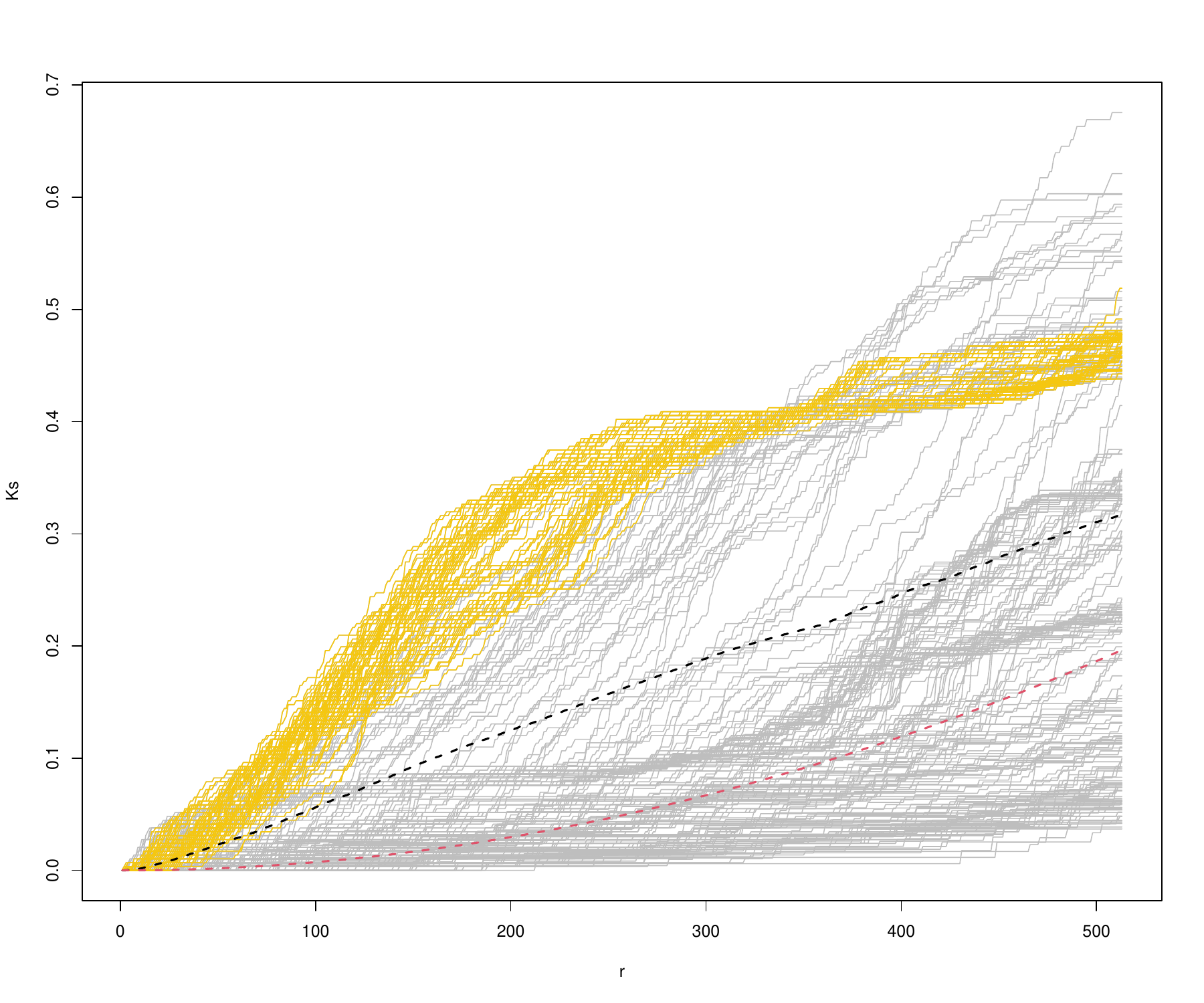}
	\caption{\textit{Left panel:} simulated clustered point pattern; \textit{Right panel:} local $K$-functions.}
	\label{fig:1a}
\end{figure}

In order to have a better view on how to choose a sensible candidate for the discrepancy $t(\cdot,\cdot)$, in Figure \ref{fig:1a} we graphically represent the estimated local $K$-functions of a simulated point pattern from a clustered point process. 
More specifically, the left panel of Figure \ref{fig:1a} shows the simulated pattern with the most dense cluster indicated in yellow; this is where we expect $\widehat\phi^*(x_i)$ to be larger than one. In the right panel, we show the corresponding local $K$-function estimates: 
in black, we have the estimated global one, i.e.\ the average of the local ones, in red, we have the theoretical one, $K_{Pois}$, and in grey, we have the estimated local ones. In yellow, we report the estimated local $K$-functions corresponding to the highlighted cluster. 
We expect larger values of $\widehat\phi^*(x_i)$ for the points in the highlighted cluster and we see that this is indeed the case.

Looking closer at Figure \ref{fig:1a}, we quickly see why the uniform metric would not be a good choice, at least not for clustered processes which should have $\phi_{\theta}^*(u)$ larger than 1 -- the largest deviation between a local $K$-function and $K_{Pois}$ is never larger than 1. Visual inspection shows that a similar argument may be applied to the $L_2$ metric. An issue that may also be present with \eqref{eq:phistar2}.
As a further option, we will also consider 
\begin{equation}
t(\hat{K}^i(r), K_{Pois}(r))
=
\exp\left\{\int_{r_0}^{r_{max}} \frac{(\hat{K}^i(r) - K_{Pois}(r))^a}{K_{Pois}(r)} \de r \right\},
\qquad x_i\in\mathbf{x},
\label{eq:phistar3}
\end{equation}
where $a=2$.

\begin{figure}[H]
    \includegraphics[width=.5\textwidth]{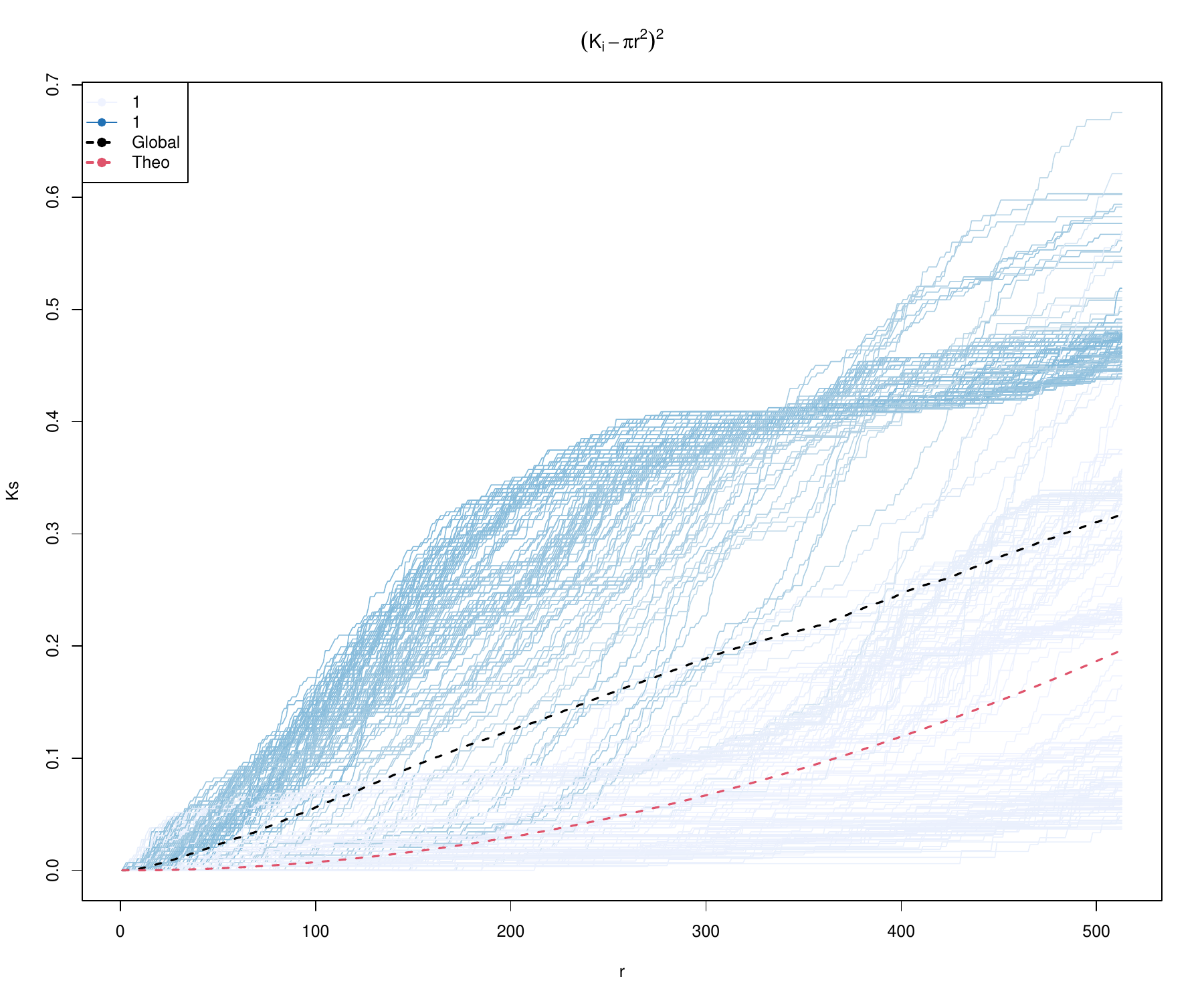}
\includegraphics[width=.5\textwidth]{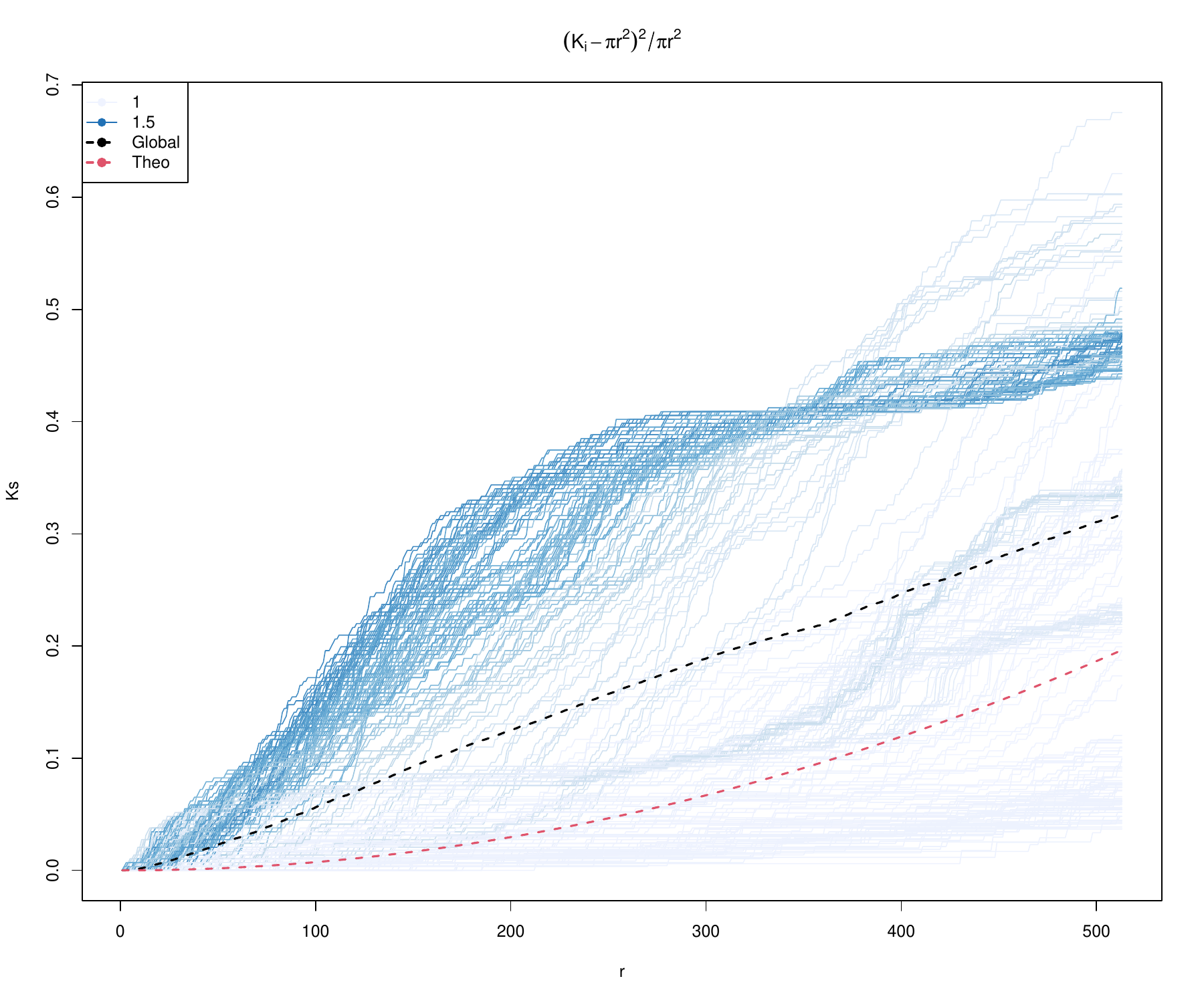}
 \caption{Local $K$-functions, coloured by the estimated $\widehat\phi^*(x_i)$ values: in the left panel, $t(\cdot,\cdot)$ is obtained through the squared difference of equation \eqref{eq:phistar2}, and in the right panel they are obtained by equation \eqref{eq:phistar3}.}
	\label{fig:2}
\end{figure}

To visually assess if a denser cluster renders higher values for $\widehat\phi^*(x_i)$, we show the graph of the local $K$-functions, coloured by the estimated $\widehat\phi^*(x_i)$ values in Figure \ref{fig:2}: the higher the value, the darker the colour. In particular, in the left panel, $t(\cdot,\cdot)$ is obtained through the squared difference of equation \eqref{eq:phistar2}, while the values in the right panel are obtained through \eqref{eq:phistar3}.

We observe that the proposal in \eqref{eq:phistar3} better succeeds in assigning higher values of $\widehat\phi^*(x_i)$ to the points of the highlighted cluster.
This motivates us to consider the proposal in equation \eqref{eq:phistar3} rather than \eqref{eq:phistar2} for our study.

\section{Simulation study}\label{sec:sims}

This section is devoted to a simulation study to assess the performance of our proposal, in terms of goodness-of-fit with respect to the intensity function. In particular, we are interested in whether there is an improvement from including the second-order characteristics in the first-order intensity function estimation, that is to say, using $\hat{\rho}_\theta(u) = \hat{\nu}_\theta(u) \hat{\phi^*}(u)$ instead of the composite likelihood choice based only on $\hat{\nu}_\theta(u)$. 
Borrowing terminology from the framework of marked point processes, we are basically comparing the first-order intensity function identified by only the ground intensity, corresponding to $\hat{\nu}_\theta(u)$, with the proposed intensity $\hat{\rho}_\theta(u)$, which weights the ground intensity by the second-order characteristics information.
We expect that the penalised likelihood leads to better fitting when considering non-Poisson processes, capturing outlying second-order behaviour present in the point pattern. 
Furthermore, regarding the interpolation used in the penalised procedure, we compare results coming from
    \begin{itemize}
        \item $\hat{\phi_\theta^*}(u)= \hat{\phi_\theta^*}(x_i)$ on observed locations, and $\hat{\phi_\theta^*}(u)= 1$ otherwise. We shall denote this as method I, which stands for \textit{indicator}. 
        Note that here \eqref{eq:CLEpen} reads 
        \begin{equation*}
        \theta\mapsto
        \sum_{x\in\mathbf{x}}
        \log({\nu}_{\theta}(x)
        \widehat\phi^*(x))
        -
        {\nu}_{\theta}(x)
        \widehat\phi^*(x)
        ,
        \end{equation*}
        by treating this case as using the smoother $B_I(u)=\sum_{x\in\mathbf{x}}\delta(u-x)$, where $\delta$ denotes the Dirac delta.
        
         \item $\hat{\phi_\theta^*}(u)$ interpolated through Inverse Distance Weighting, i.e.\ $B_{idw}(u)$, where the power is set to $p=2$; this case will be denoted by IDW.
          \item $\hat{\phi_\theta^*}(u)$ interpolated by means of kernel smoothing, i.e.\ $B_{kernel}(u)$; this case will be denoted by KS.
    \end{itemize}

Let $\rho(\cdot)$ denote the intensity function of the generating process, $X$, henceforth called the \textit{true} intensity function. 
Given an intensity estimator $\hat{\rho}(\cdot) = \hat{\rho}(\cdot;X)$, when the true intensity is known, the goodness-of-fit of $\hat{\rho}(\cdot)$ may be measured by the \textit{mean integrated squared error}, 
$$
\mathrm{MISE}_{\hat{\rho}} = \mathbb{E}\Biggl[\int_W\big(\hat{\rho}(x) - \rho(x)\big)^2dx\Biggl]
=
\int_W\mathrm{Var}\left(\hat{\rho}(x)\right)dx
+
\int_W\mathrm{Bias}(\hat{\rho}(x))^2dx
,
$$
where $\mathrm{Bias}(\hat{\rho}(x)) = \mathbb{E}[\hat{\rho}(x)] - \rho(x)$. 
Since it is hard to compute $\mathrm{MISE}_{\hat{\rho}}$ analytically for a given intensity estimator $\hat{\rho}$, we resort to simulations and estimate $\mathrm{MISE}_{\hat{\rho}}$ by $\frac{1}{n}\sum_{i=1}^n \int_W\big(\hat{\rho}(x;\mathbf{x}_i) - \rho(x)\big)^2dx$, given realisations $\mathbf{x}_1,\ldots,\mathbf{x}_n$ of $X\subseteq W$.

If, however, the true intensity function is not known explicitly, which is actually the common situation (recall the discussion after expression \eqref{eq:PapangelouInt}), 
we instead employ the Pearson $\chi^2$ statistic based on quadrat counts to evaluate the goodness-of-fit. 
Here, the window of observation $W$ is divided into disjoint tiles, $W_i$, $i=1,\ldots,m$, $\bigcup^m_{i=1} W_i
= W$, and the number of data points in each tile is counted. The tiles are usually rectangular, but may have arbitrary shapes. Then, the expected number of points in each quadrat is computed, as determined by the fitted model. 
Note that for a fitted intensity $\hat{\rho}(u)$, $u\in W$, the predicted number of points falling in any $W_i$ is $\int_{W_i} \hat{\rho}(u)du$. 
More specifically, the resulting Pearson $\chi^2$ statistic is given by
\begin{equation}
\chi^2_{\hat{\rho}} = \sum_{i = 1}^{m}
\frac{(n(\mathbf{x} \cap W_i) - \int_{W_i} \hat{\rho}(u)du)^2}{\int_{W_i} \hat{\rho}(u)du},
    \label{eq:chi}
\end{equation}
where $n(\mathbf{x} \cap W_i)$ denotes the number of points of the observed point pattern $\mathbf{x}$ in $W_i$. 
In our simulation study, $\chi^2_{\hat{\rho}}$ will be computed for a large number of simulated realisations $\mathbf{x}$ of the generating process $X$, and the mean of these values will be the value which determines the fit; the lower the value, the better the fit of an estimator $\hat{\rho}$.

\subsection{Models with tractable intensity functions}
We consider the following three model families, which are of particular interest because we actually know the form of $\phi_{\theta}^*$ explicitly:
\begin{itemize}
    \item Poisson processes. Here, $\phi_{\theta}^*(u)=1$.
    \item Log-Gaussian Cox processes (LGCPs). Let the driving random field $\Lambda_{\theta}(u)={\nu}_{\theta}(u)\exp\{Z(u)\}$ be such that the zero mean stationary Gaussian random field $Z$ has covariance function $C$ satisfying $C(u,u)=\sigma^2$. Then, $\phi_{\theta}^*(u)=\exp\{\sigma^2/2\}$.
    \item Determinantal point processes (DPPs). Let $\bar\rho_{\theta}=\sup_{u\in W} {\nu}_{\theta}(u)$ and consider a homogeneous DPP with kernel given by a stationary covariance function satisfying $C(u,u)=\bar\rho_{\theta}\sigma^2$; this is also the intensity of the homogeneous DPP.  Then, by independently thinning it with retention probability $p_{\theta}(x,y)={\nu}_{\theta}(u)/\bar\rho_{\theta}\in[0,1]$, $u\in W$, we obtain an inhomogeneous DPP with $\phi_{\theta}^*(\cdot)=\sigma^2$ and intensity function 
    $$
\rho_{\theta}(x,y)
=
p_{\theta}(x,y)
\bar\rho_{\theta} \sigma^2
=
{\nu}_{\theta}(u)
\sigma^2
=
\exp\{\theta_0 + \theta_1 z_1(u) + \cdots + \theta_d z_d(u)\}\sigma^2.
$$  
\end{itemize}

Throughout, we will restrict ourselves to the spatial domain $W=[0,1]^2$. The specific model choices provided below are taken from \citet{cronie2018non, moradi2019resample}.

\subsubsection{Poisson processes}

In the first scenario, we consider a homogeneous Poisson process with constant intensity $\rho \in \{125, 250, 500\}$, directly representing the expected number of points in $W=[0,1]^2$. 

Then, we consider an inhomogeneous Poisson process, with a linear intensity function of the form
$$
\rho(x,y)= 10 + \alpha x, \qquad
(x,y) \in [0,1]^2,
$$ 
where we let $\alpha \in \{240,480,960\}$, leading again to having $\{125, 250, 500\}$ as expected numbers of points.

Finally, we simulated from an inhomogeneous Poisson process with modulation, that is, one with intensity function 
$$
\rho(x,y)=\alpha + \beta \cos{(10 x)}, \qquad
(x,y) \in [0,1]^2.
$$  
We take $\beta=100$ and $\alpha\in\{125, 250, 500\}$.

\subsubsection{Log-Gaussian Cox processes}

The next model family on which we evaluate our intensity estimation approach is LGCPs. 
These are generally used to describe and analyse aggregated phenomena.
Specifically, let $Z$ be a Gaussian random field with mean zero and covariance
 $$
 \sigma^2 \exp{(-\beta||(x_1,y_1) - (x_2,y_2)||)},$$
 where $\sigma^2$ is the variance and $\beta$ is the scale parameter for the spatial distance. The exponential form is widely used in this context and nicely reflects the decaying correlation structure with distance.
Consider further the random field defined by 
$\rho \exp{Z(x,y)}$, which has constant mean function.
Then, the intensity function of the resulting Cox process, a homogeneous LGCP, is given by
$\rho\exp\{\sigma^2/2\}$.

The first scenario here involves fixing $\sigma^2$ and $\beta$ to $0.15$ and $0.5$, respectively, and letting $\rho$ be the logarithm of the expected number of points to simulate.

Shifting to the more critical scenario of inhomogeneity combined with clustering, consider the random field 
$\rho(x,y)\exp{(Z(x,y))}$, with $\rho(\cdot)$ strictly positive. Then, the resulting Cox process has intensity function $\rho(x,y)\exp{(\sigma^2/2)}$. 
Here, we let
$$\rho(x,y) = \alpha - 1.5  (x - 0.5)^2 + 2  (y - 0.5)^2,$$ where $\alpha \in \{\log{(125)}, \log{(250)}, \log{(500)}\}$ controls the expected number of points.
Furthermore, we also consider a homogeneous LGCP with $\rho = \log{45}$,  $\sigma= 5$, and $\beta= 0.05$, which exhibits a more clustered behaviour.

\subsubsection{Determinantal point processes} 

Finally, we considered the regular/inhibitory model family given by DPPs. Here, the occurrence of an event
has an inhibiting effect on the occurrence of having close by neighbours.

DPPs 
allow explicit expressions for the product densities
$$
\rho^{(n)}((x_1,y_1), \ldots, (x_n,y_n))=\det{(C((x_i,y_i),(x_j,y_j)))_{i,j}}, 
\qquad (x_i,y_i) \in W,\quad n\geq1,
$$
in terms of the determinant of a matrix generated by a general kernel $C$, typically taken to be a covariance function \citep{lavancier2015determinantal}. Hereby, the intensity function is given by $\rho(x,y)=\rho^{(1)}(x,y)=C((x,y),(x,y))$.
In this paper, we let 
$$C((x_1,y_1),(x_2,y_2))=\sigma^2\exp{(-\beta||(x_1,y_1)-(x_2,y_2)||)},$$ resulting in a homogeneous DPP with intensity $\rho(x,y)=C((x,y),(x,y))=\sigma^2.$

Here, $\beta$ is taken as 50, and $\sigma$ plays the role of the intensity, which we let take values in $\{125,250\}$.

To combine inhomogeneity and repulsive behaviour of the points, we apply independent thinning to the realisations of the homogeneous DPP just described. Choosing some retention probability $p(x,y)$, $(x,y) \in [0,1]^2$, results in the intensity function $\rho(x,y)=\sigma^2p(x,y)$. 
In this scenario, we choose the retention probability function 
$$
p(x,y) = \frac{10 + 80x}{90}, \qquad (x,y) \in [0,1]^2.
$$

\subsubsection{Numerical results}

For each of the indicated models, we generated 100 independent realisations on the unit square, with different expected number of points $\approx \mathbb{E}[N]$,
in order to 
assess the goodness-of-fit based on $\mathrm{MISE}$. For each specific model, we compared the proposed penalised Poisson likelihood to the Poisson log-likelihood, where $\phi_\theta^*(u)=1$. 
The results can be found in Table \ref{tab:2bis}; note that the smaller the value of $\mathrm{MISE}$, the better the fit.


\begin{table}[H]
    \centering
\caption{Results of the simulation study: MISE over 100 simulations}
\begin{tabular}{l|l|r|rrr}
\toprule
Scenario  &$\mathbb{E}[N]$&$\hat{\nu}_\theta(x_i)$  &\multicolumn{3}{c}{$\hat{\rho}_\theta(x_i)$} \\ 
       & &   &I& IDW &  KS  \\ 
\midrule
Poisson  &125& 2093220& 2093113 &2093624& 2093763   \\ 
(Homogeneous)  &250&   3815670& 3816500 &3816099 &3816261  \\ 
  &500&  8249836& 8246385& 8250239& 8250443 \\ [.1cm]
Poisson  &125& 658 &656 &666& 668\\ 
(Inhomogeneous)  &250& 2010 &2001& 2040& 2048 \\ 
  &500&  6598 &6567 &6730& 6772\\ [.1cm]
 Poisson    &125& 5218  & 5217& 277435 &277545\\
(Modulated)  &250& 5424  & 5423 &196459 &196567 \\ 
  &500& 5918   &5918& 128257& 128361 \\ \hline
LGCP     &125& 1960 &1960 &1956 &1956 \\ 
(Homogeneous)  &250& 8490 &8490& 8463& 8464\\ 
  &500& 29277 &29277 &29218 &29222\\ 
LGCP     &125 &2972 &2971& 2962& 2963\\ 
 (Inhomogeneous) &250& 13266& 13265 &13212 &13212\\ 
  &500& 50332& 50332& 50134& 50129 \\ [.1cm]
  LGCP    &125& 1190807& 1178058 &1186114 &1183964 \\ 
(Homogeneous,  &250&5263696 &5131765& 5241202 &5236869\\ 
 more clustered)  &500& 31885099 &29259048 &31695509 &31682127\\ \hline
DPP&125&  15873 &15870 &15873 &15873\\ 
  (Homogeneous)&250& 63030 &63022& 63030 &63030\\ [.1cm]
DPP     &125&  5019& 5017 &5019& 5019\\ 
 (Inhomogeneous) &250&  19611 &19602& 19611 &19611 \\ 
\bottomrule
\end{tabular}
    \label{tab:2bis}
\end{table}

By examining Table \ref{tab:2bis}, it appears that for most scenarios, the penalised Poisson likelihood (method I without interpolation) generally produces the lowest $\mathrm{MISE}$ values, compared to the other methods. 
In particular, the kernel smoothing interpolation method (KS) tends to have higher MISE values, suggesting it might not be the most accurate method for intensity estimation in these scenarios, while the IDW method shows varying performance across scenarios, indicating its sensitivity to the underlying data distribution. 
Therefore, albeit slightly, our proposed method seems to outperform the rest of the considered approaches. In particular, the results indicate that there is merit to the type of local dependence scaling we propose here. 

\subsection{Models with intractable intensity functions}

On the basis of the results above, we further proceed by simulating more complex scenarios, where points tend to cluster around or inhibit each other as a result of particular second-order structures. More specifically, we consider Thomas and Strauss processes.

\subsubsection{Thomas processes}
For a Thomas process \citep{thomas49, waagepetersen07}, let  $\kappa$ be the intensity of the Poisson process of cluster centres, $\sigma$ the standard deviation of a random displacement (along each coordinate axis) of a point from its cluster centre, and $\mu$ the mean number of points per cluster.
In the simplest case, where $\kappa$ and $\mu$ are scalars, a uniform Poisson point process of \textit{parent} points with intensity $\kappa$ is generated. Then, each parent point is replaced by a random cluster of \textit{offspring} points, the number of points per cluster is Poisson distributed with mean $\mu$, and their positions are isotropic Gaussian displacements from the cluster parent location. The resulting point pattern is a realisation of the classical \textit{stationary Thomas process} generated inside the window $W$, with intensity $\kappa \mu$.
Inhomogeneous versions of the Thomas process imply the intensity function of an inhomogeneous Poisson process that generates the parent points.

\subsubsection{Strauss processes}
The Strauss process \citep{strauss75, ripley-kelly76, ripley-kelly77}, is a model for spatial inhibition, that depends on the value of the interaction parameter $\gamma\leq 1$ and the interaction radius 
$R$, such that each pair of points closer than 
$R$ units apart contributes a factor 
$\gamma$ to the density.    This parameter regularises the \textit{ordered} or \textit{inhibitive} effect of the pattern: if $\gamma=1$, the Strauss process reduces to a Poisson process; if $\gamma=0$, the Strauss process is called hard core process with hard core radius 
$R/2$, since no pair of points can lie closer than $R$ units apart.

\subsubsection{Numerical results}

In this case, we generate 100 realisations from each of the following six scenarios:
\begin{enumerate}
 \item \label{scen:2} Thomas process with $\kappa = 20$ intensity of the Poisson process of cluster centers and  $\mu(x, y) = 5  \exp(2 x - 1) $ points constituting each cluster in a disc of radius 0.2. 
    \item \label{scen:3} Thomas process with $\kappa = 25$, $\mu(x, y) = 5  \exp(2 x - 1) $ and radius 0.2. 
   \item \label{scen:4} Thomas process with $\kappa = 50$, $\mu(x, y) = 5  \exp(2 x - 1) $ and  radius 0.2. 
       \item \label{scen:7} Strauss process with interaction parameter $\gamma= 0.3$ and interaction radius 0.05. 
   \item \label{scen:8} Strauss process with interaction parameter $\gamma= 0.5$ and interaction radius 0.05. 
   \item \label{scen:10} Strauss process with interaction parameter $\gamma= 0.7$ and interaction radius 0.05. 
\end{enumerate}

\begin{figure}[H]
    \centering
    \includegraphics[width = .475\textwidth]{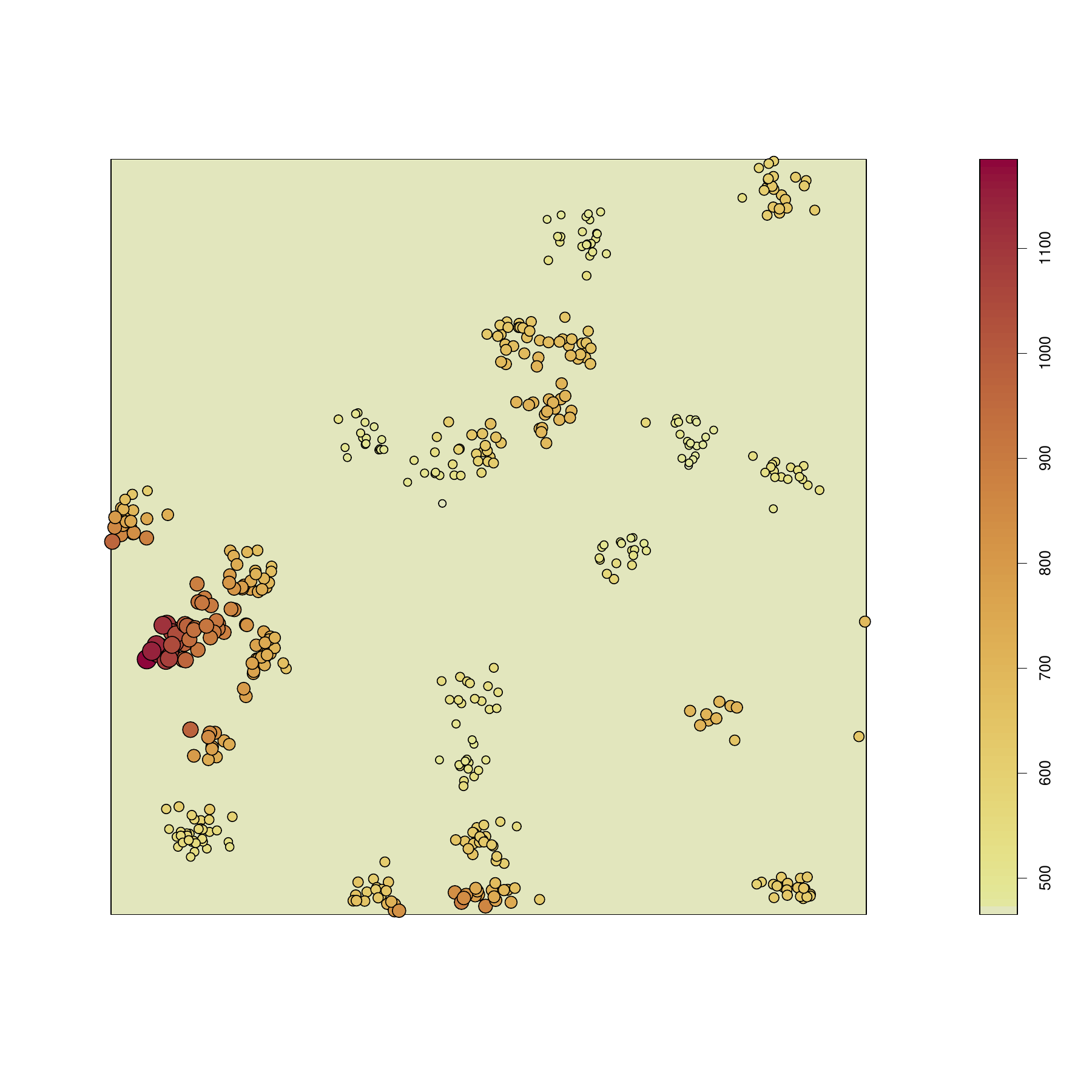}
    \includegraphics[width = .475\textwidth]{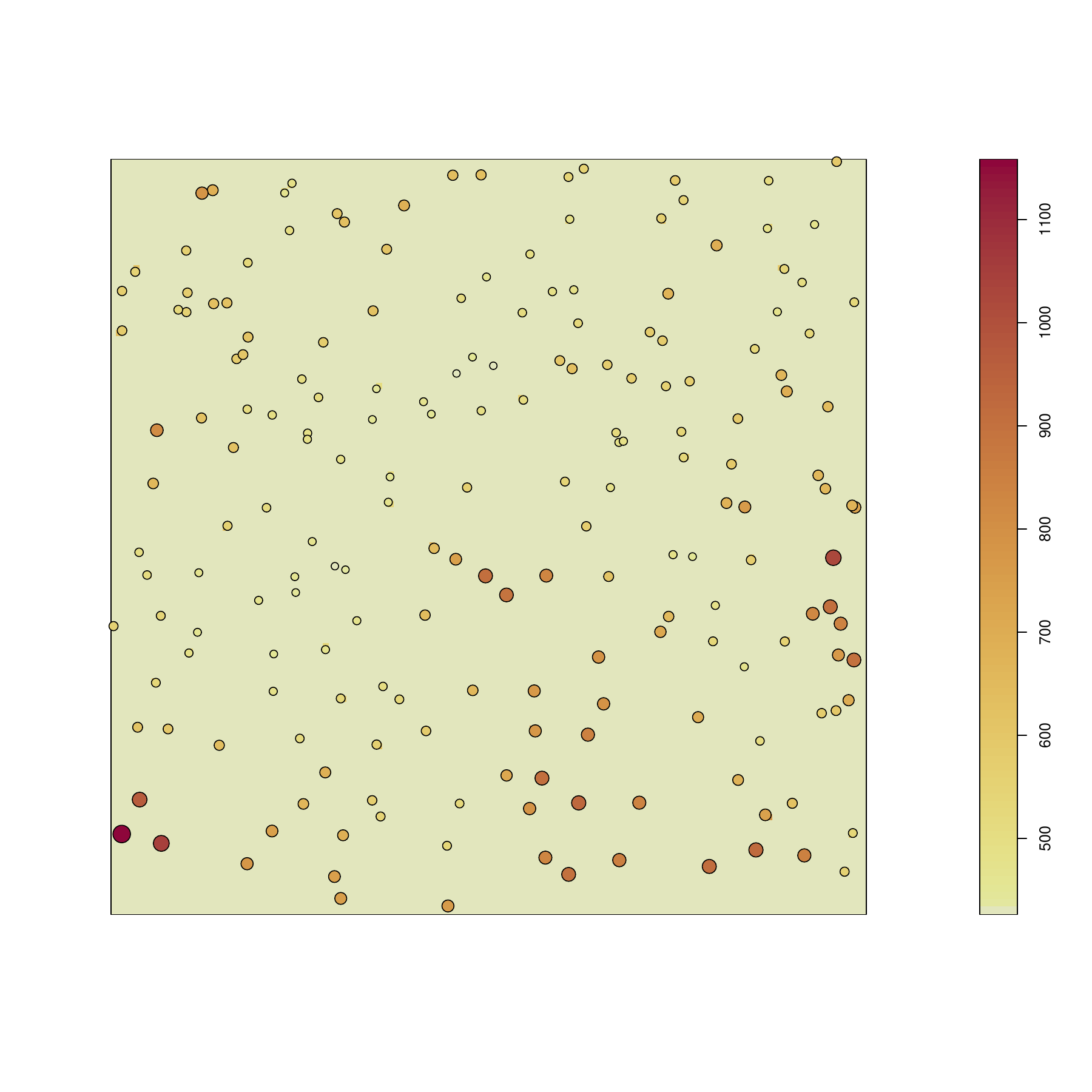}
    \caption{Simulated patterns with the superimposed intensity fitted by means of the proposed penalised likelihood. (a) Thomas process following scenario \ref{scen:2}; (b) Strauss process following scenario \ref{scen:10}. 
    }
    \label{fig:1}
\end{figure}

Figure \ref{fig:1} shows two simulated patterns from the Thomas process of scenario \ref{scen:2}, and the Strauss process of scenario \ref{scen:10}, and with the superimposed intensity fitted by means of the proposed penalised likelihood. 

As the intensity is not known in explicit form, the Pearson $\chi^2$ statistics based on quadrat counts can be employed for assessing the goodness-of-fit. 
Table \ref{tab:1bis} reports the means 
over 100 simulations of the    $\chi^2$-statistics, for both the intensities  $\hat{\nu}_\theta$ and $\hat{\rho}_\theta$, for the different scenarios. Here, a lower value indicates a better fit.

\begin{table}[H]
    \centering
\caption{Results of the simulation study: $\chi^2$ over 100 simulations}
\begin{tabular}{l|l|r|rrr}
\toprule
Scenario  &$\mathbb{E}[N]$&$\hat{\nu}_\theta(x_i)$&&$\hat{\rho}_\theta(x_i)$& \\ 
       & &  &I& IDW &  KS  \\ 
\midrule
Thomas  [\ref{scen:2}] &115& 15273 &15241&15287&15297 \\ 
Thomas [\ref{scen:3}]&150&25567 &25476 &25522&25538\\ 
Thomas  [\ref{scen:4}]& 300&  89312 &89127&89202&89189\\  
\midrule
Strauss [\ref{scen:7}]&120&   16323.32 &16323.29&16323.53&16323.57\\ 
Strauss [\ref{scen:8}]& 200&16314.38 &16314.40&16314.44&16314.45\\ 
Strauss [\ref{scen:10}] & 400&16330.5 &16330.5&16330.52&16330.52\\ 
\bottomrule
\end{tabular}
    \label{tab:1bis}
\end{table}

The results in Table \ref{tab:1bis} indicate that also here, where there are strong interactions,  $\hat{\rho}_\theta$ improves the fit with respect to $\hat{\nu}_\theta$.  
According to the $\chi^2$-statistic for the Thomas scenarios, our proposal seems to perform better than the Poisson process log-likelihood. However, 
 this improvement can not be said to be confirmed for the Strauss scenarios, showing only slightly lower values in the penalised case. This is likely due to the difficulties in spotting the repulsive behaviour of patterns with smaller sizes, wrongly identified as Poisson ones.

\section{Application to the Redwood clustered data}\label{sec:appl}

In this section, we show a practical application of our proposal by analysing a spatial pattern of the Redwood data available in the \texttt{spatstat} package \citep{spatstat}. 

The aim is first to compare our proposal to a model for clustered patterns, and second, to show the implications of adding the penalisation to a Poisson likelihood.

The data represent the locations of 195 seedlings and saplings of California Giant Redwood (Sequoiadendron giganteum) in a square sampling region.

\begin{figure}[H]
    \centering
    \includegraphics[width = .475\textwidth]{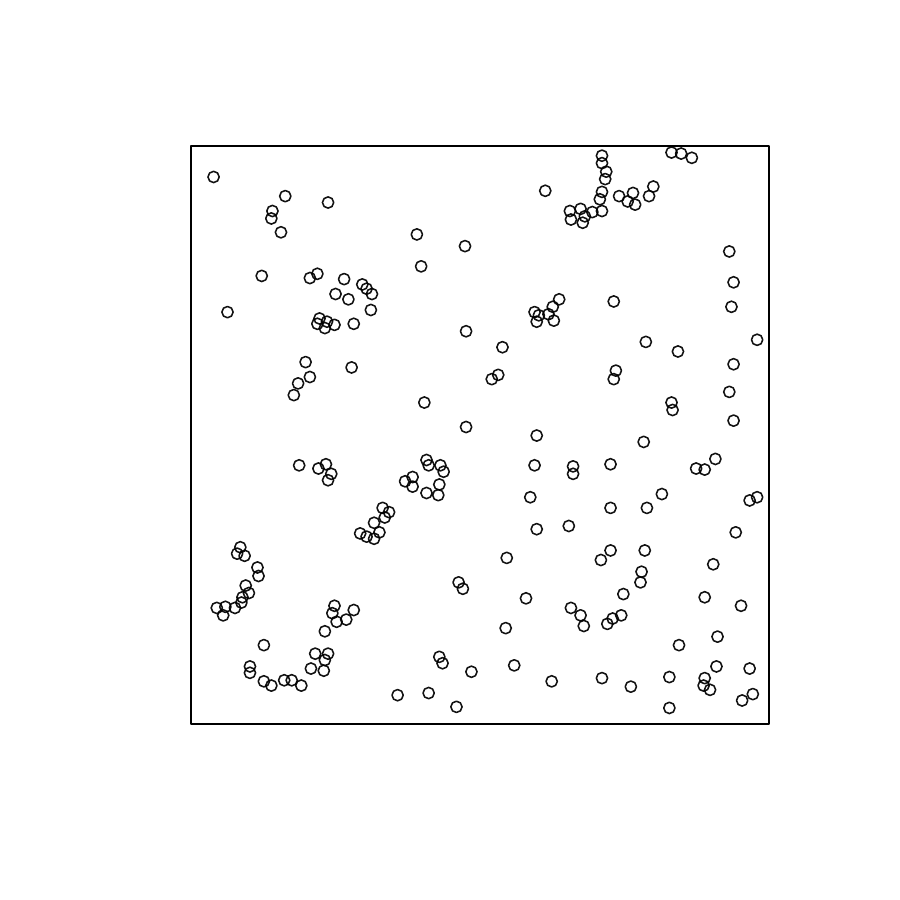}
        \includegraphics[width = .475\textwidth]{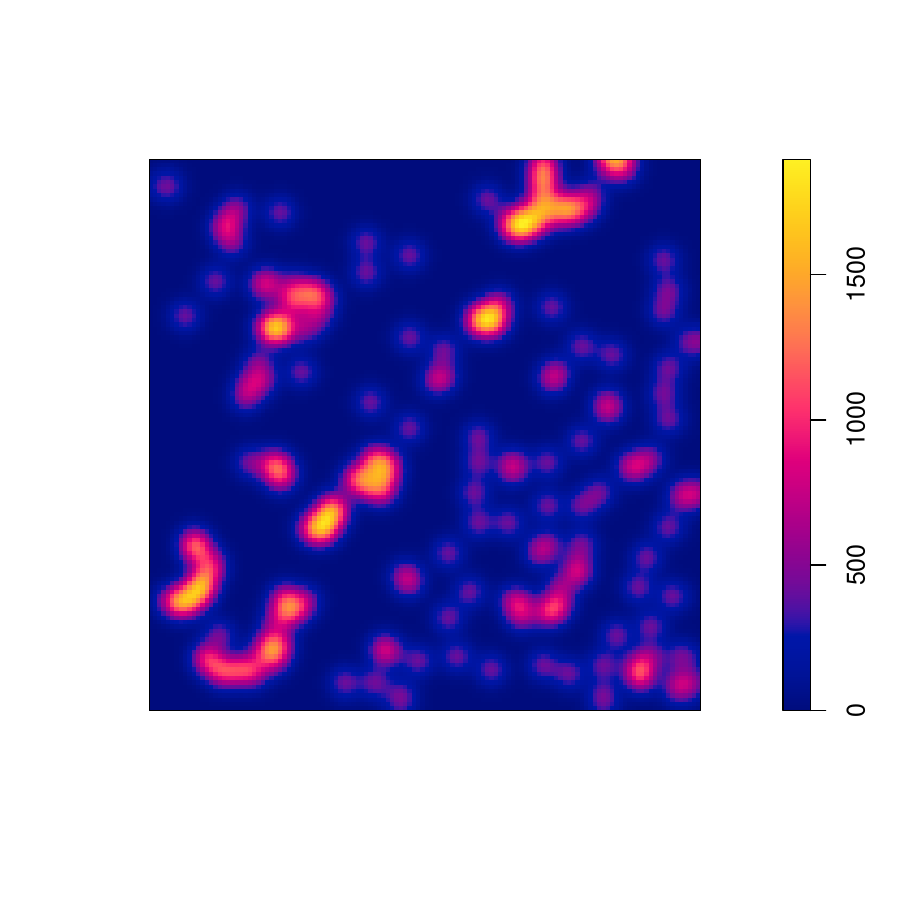}
    \caption{\textit{Left panel}: the Redwood data. \textit{Right panel}: a kernel fitted intensity function.}
    \label{fig:22}
\end{figure}

They were described and analysed by \cite{strauss75}, who divided the sampling region into two subregions: the spatial pattern appears to be strongly clustered in the upper left region and slightly regular in the remaining, as shown in the left panel of Figure \ref{fig:22}.

In the right panel of Figure \ref{fig:22}, we display $\tilde{\rho}({u}) 
$, that is, a non-parametric estimate of the intensity fitted through a kernel procedure. 
The smoothing bandwidth is selected by cross-validation as the value that minimises the Mean Squared Error criterion defined by \cite{diggle1985kernel}, by the method of \cite{berman1989estimating}. This is to consider an adaptive method and, therefore, to resemble the true intensity function the most. 

\begin{figure}[H]
    \centering
    \includegraphics[width = .9\textwidth]{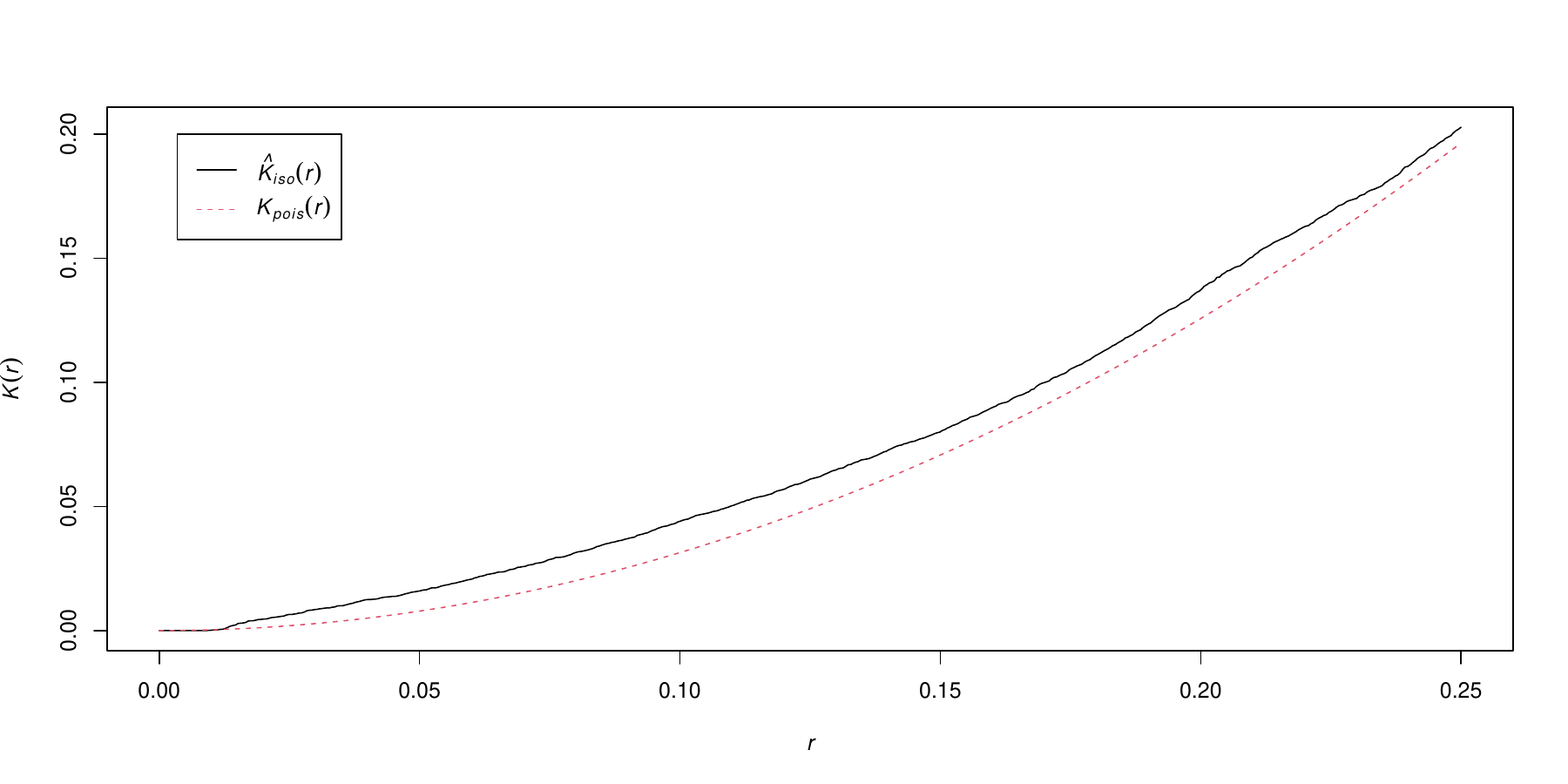}
    \caption{Estimated global $K$-function of the Redwood point pattern}
    \label{fig:6}
\end{figure}

Looking at the estimated global $K$-function of the analysed pattern, this clearly indicates clustering (Figure \ref{fig:6}). For this reason, a point process model tailored for clustering behaviour could be more appropriate.
A possible candidate is the previously introduced Thomas process.  


Proceeding by fitting such a model, fitted cluster parameters are estimated to be 82.45 and 0.026, with a mean cluster size of 2.37 points.
Despite modelling a clustered point pattern with a point process model, the fitted spatial intensity relies only on the specification of the first-order intensity function. Consequently, this results in a constant intensity similar to the unpenalized Poisson case.

Although the cluster parameters fitted by the Thomas model offer valuable insights into the clustering degree of the analysed pattern, our proposed approach facilitates this understanding by the definition of the offset. Most importantly, the proposed method enables us to account for this aspect prior to model fitting.

\begin{figure}[H]
    \centering
    \includegraphics[width = .475\textwidth]{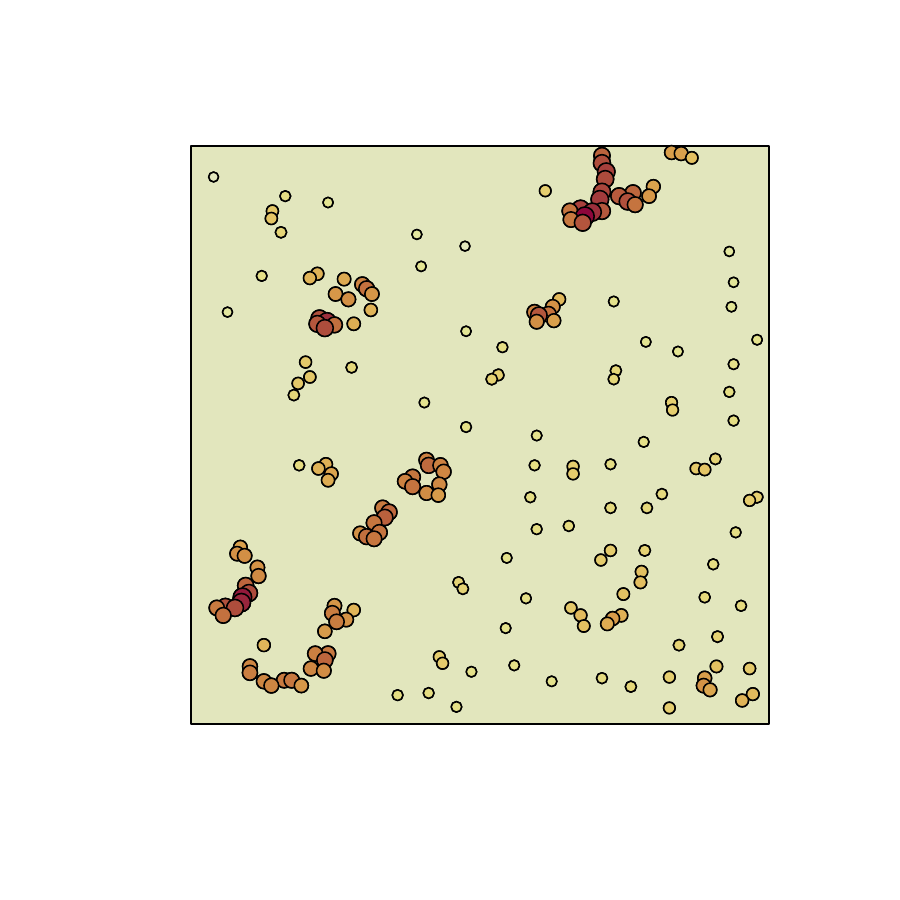}
        \includegraphics[width = .45\textwidth]{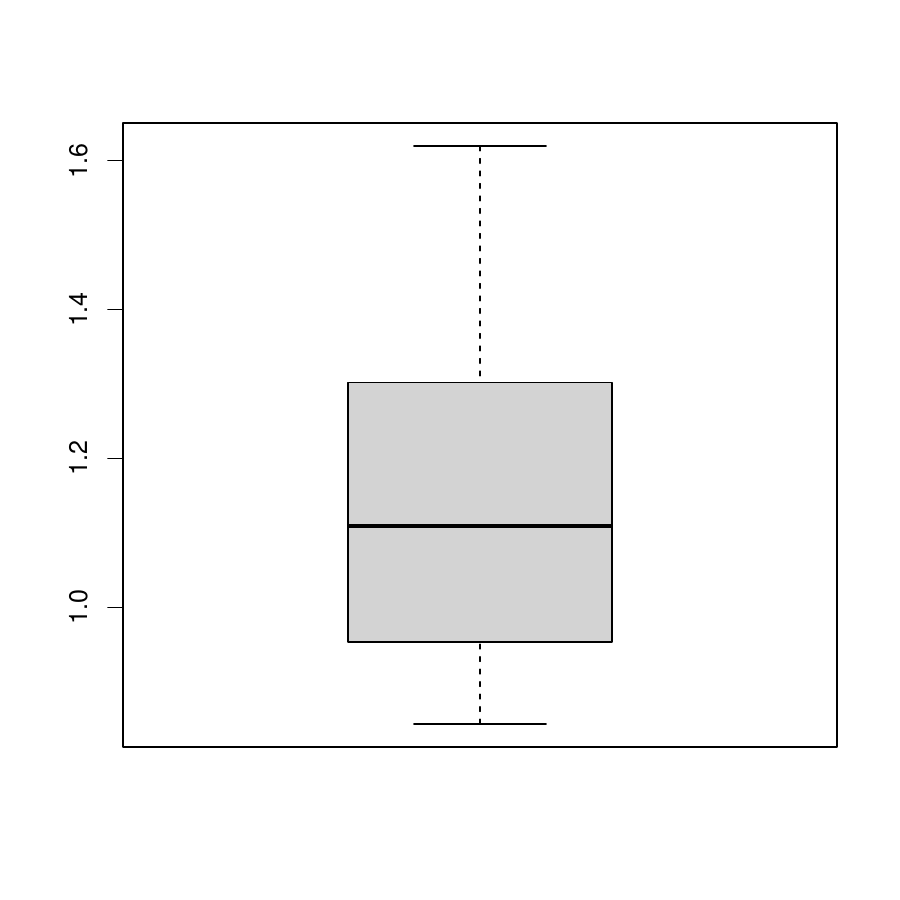}
    \caption{$\hat{\phi^*}(x_i)$ values of the Redwood pattern. \textit{Left panel}: $\hat{\phi^*}(x_i)$ values spatially displayed. \textit{Right panel}: Boxplot of the $\hat{\phi^*}(x_i)$ values.}
    \label{fig:8}
\end{figure}

 Figure \ref{fig:8} shows the $\hat{\phi^*}(x_i)$ values only at the point locations $x_i$ of the analysed Redwood spatial point pattern.

Both the spatial display of such values and their distribution indicate an overall clustering behaviour of the observed pattern, with the majority of the points exhibiting $\hat{\phi^*}(x_i)$ values greater than $1$.
In addition, this representation also enables us to pinpoint the points whose local structure is evidently more extreme than the others.

We emphasize the significance of these results, as they allow us to quantify the local clustering behaviour of the points patterns and account for it in the model fitting. Most importantly, this has been achieved without making any assumptions about the underlying process nature.

\begin{figure}[H]
    \centering
    \includegraphics[width = .475\textwidth]{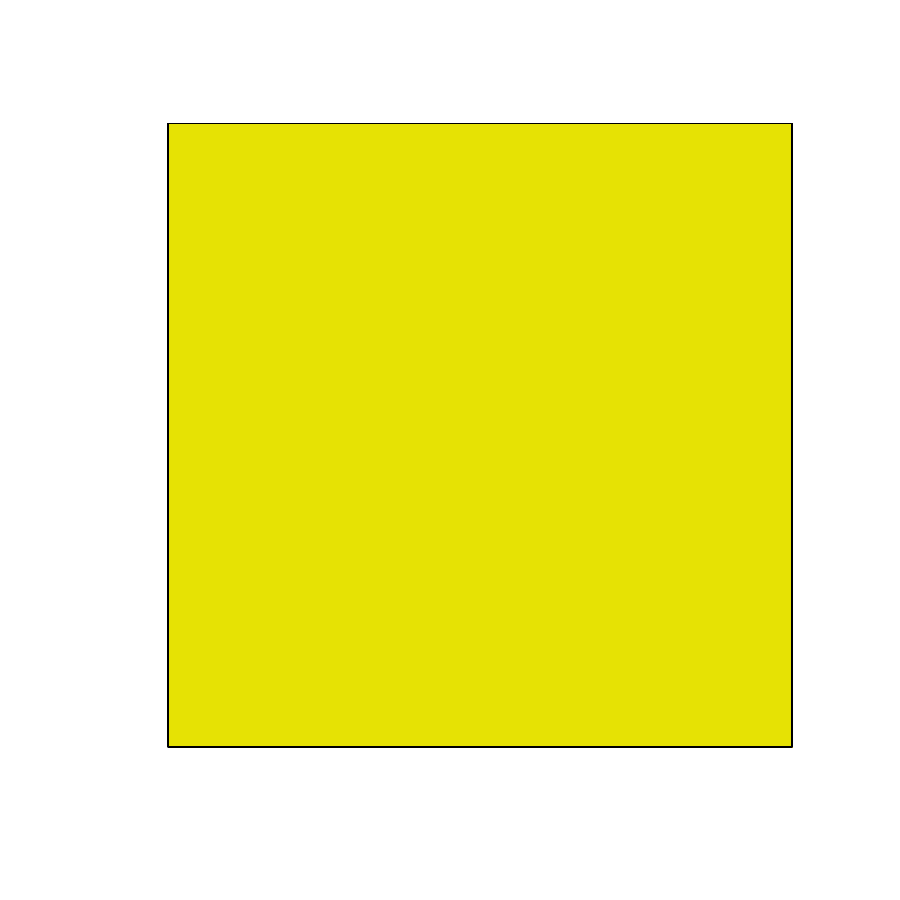}
    \includegraphics[width = .475\textwidth]{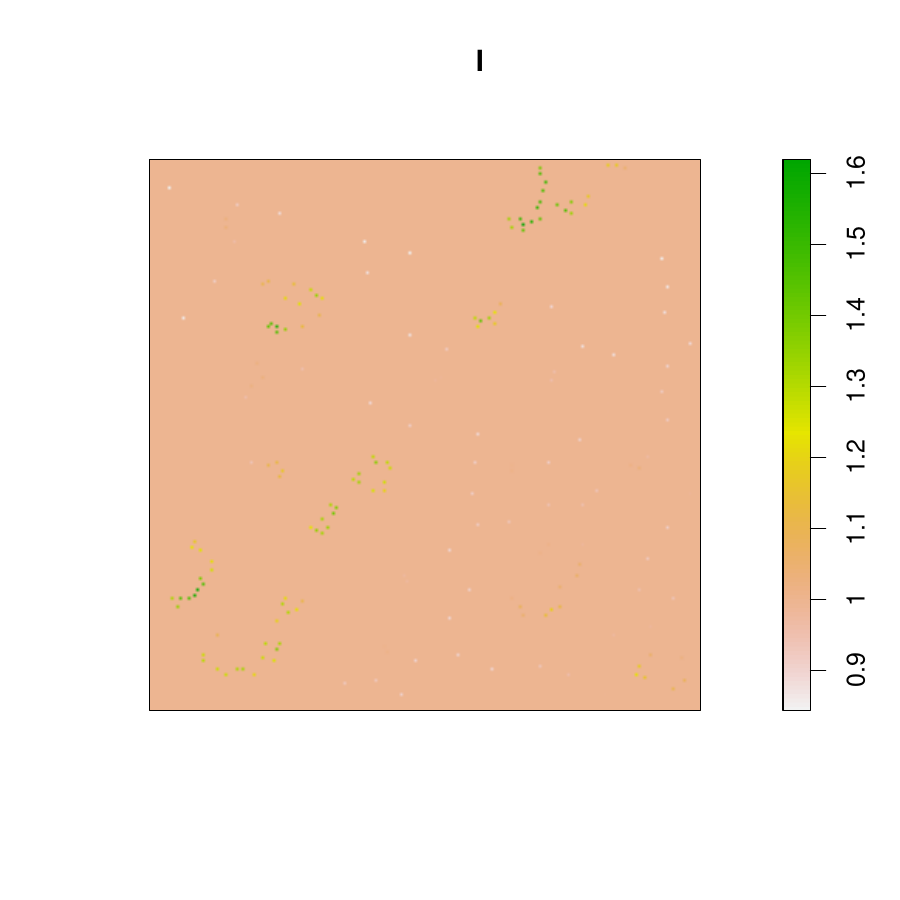}\\
        \includegraphics[width = .475\textwidth]{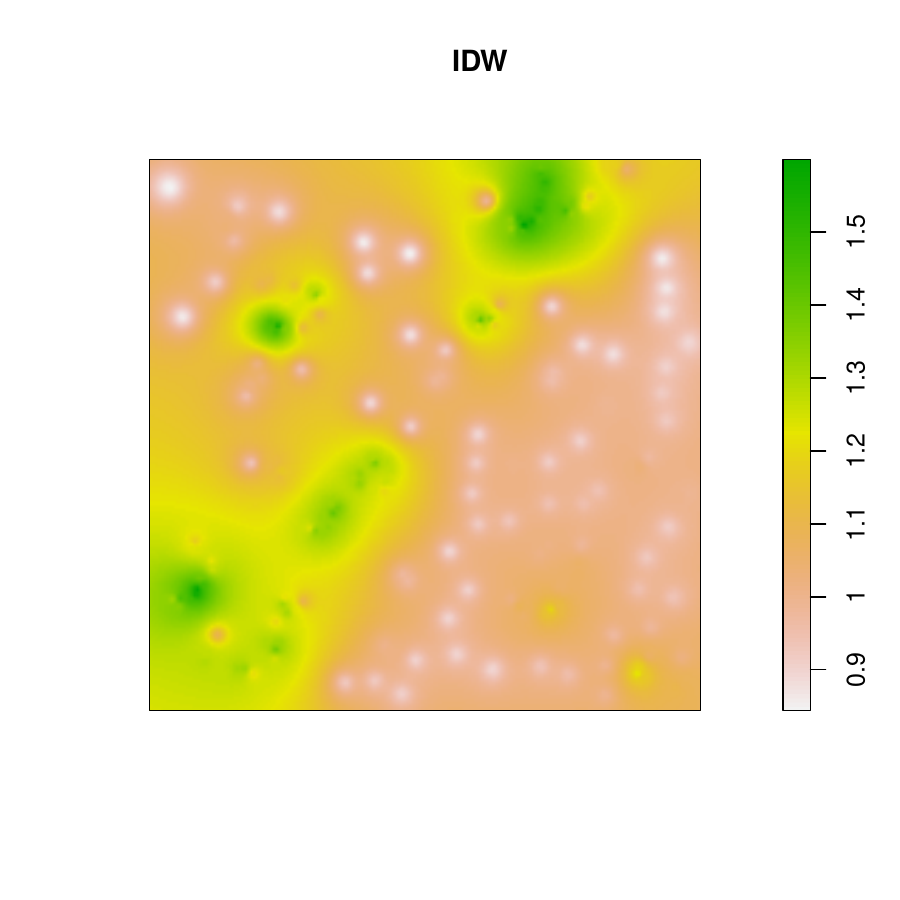}
    \includegraphics[width = .475\textwidth]{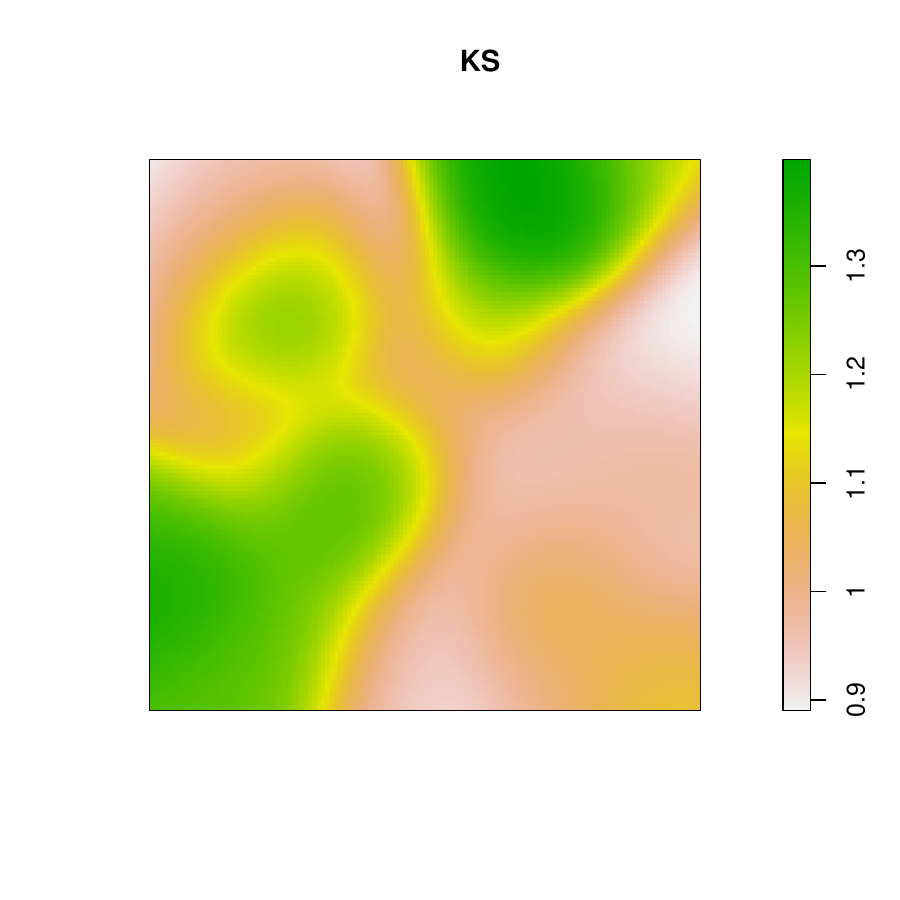}
    \caption{The offsets $B(u)=\widehat\phi^*(u)$, $u\in W$, for (a) the unpenalised case,  (b) the \textit{indicator} method, (c) the IDW methods, and (d) the kernel smoothing method.}
    \label{fig:3}
\end{figure}

Next, in order to assess the performance of the proposed penalised methods, we first compute the possible offsets $B(u)=\widehat\phi^*(u)$, $u\in W$ to include in the linear predictor of Poisson models. These are shown in Figure \ref{fig:3}, and correspond to (a) the unpenalised case, that is, $\hat{\phi^*}(u)= \hat{\phi^*}(x_i)$ on observed locations, and $\hat{\phi^*}(u)= 1$ otherwise, (b) the \textit{indicator} method, (c) the IDW method, and (d) the kernel smoothing method, introduced in the simulation Section \ref{sec:sims}.

First, we proceed by fitting four different homogeneous point process models,  using the four offsets as described above. Table \ref{tab:appl} contains the obtained AIC values for the fitted Poisson models. The results show that the fitting method of the model improves when moving to a penalised fitting, with the largest improvement for the \textit{indicator} method.

\begin{table}[h]
\centering
\caption{AIC values of the four Poisson models fitted to the Redwood data}
\begin{tabular}{|c|cccc|}
\toprule
\textbf{Offset} & - & I & IDW & KS \\
\midrule
\textbf{AIC} & -1664.47 & -1713.779 & -1673.381 & -1678.455 \\
\bottomrule
\end{tabular}
\label{tab:appl}
\end{table}

\begin{figure}[H]
    \centering
    \includegraphics[width = .475\textwidth]{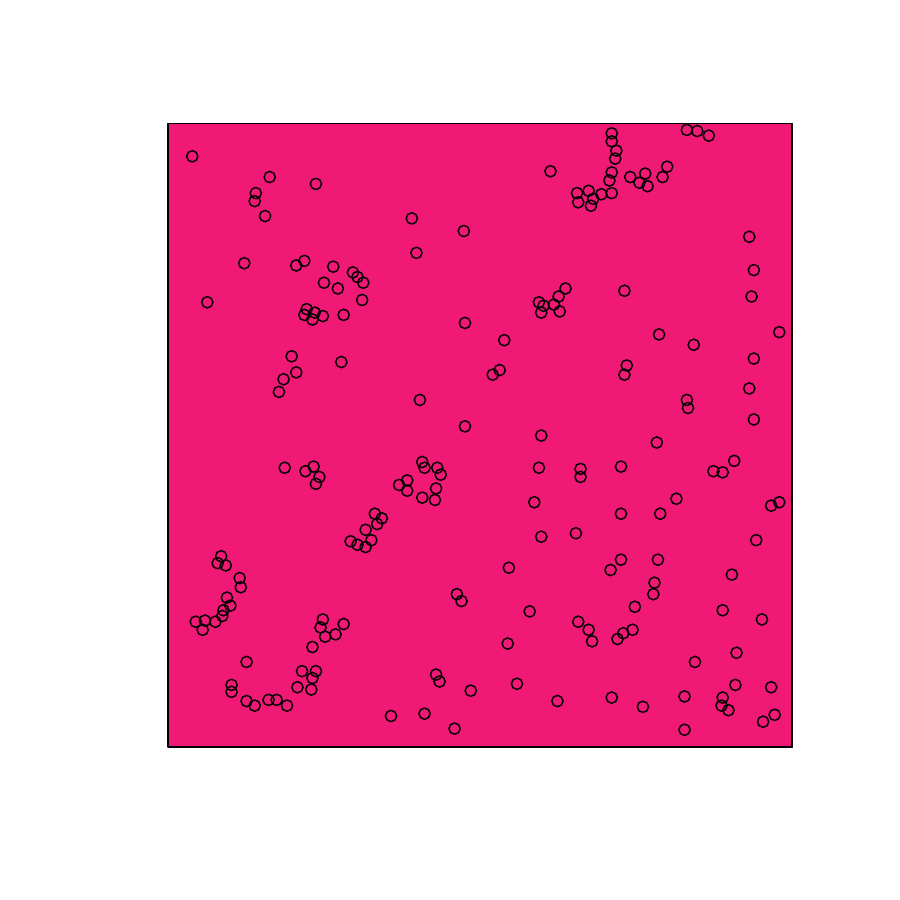}
    \includegraphics[width = .475\textwidth]{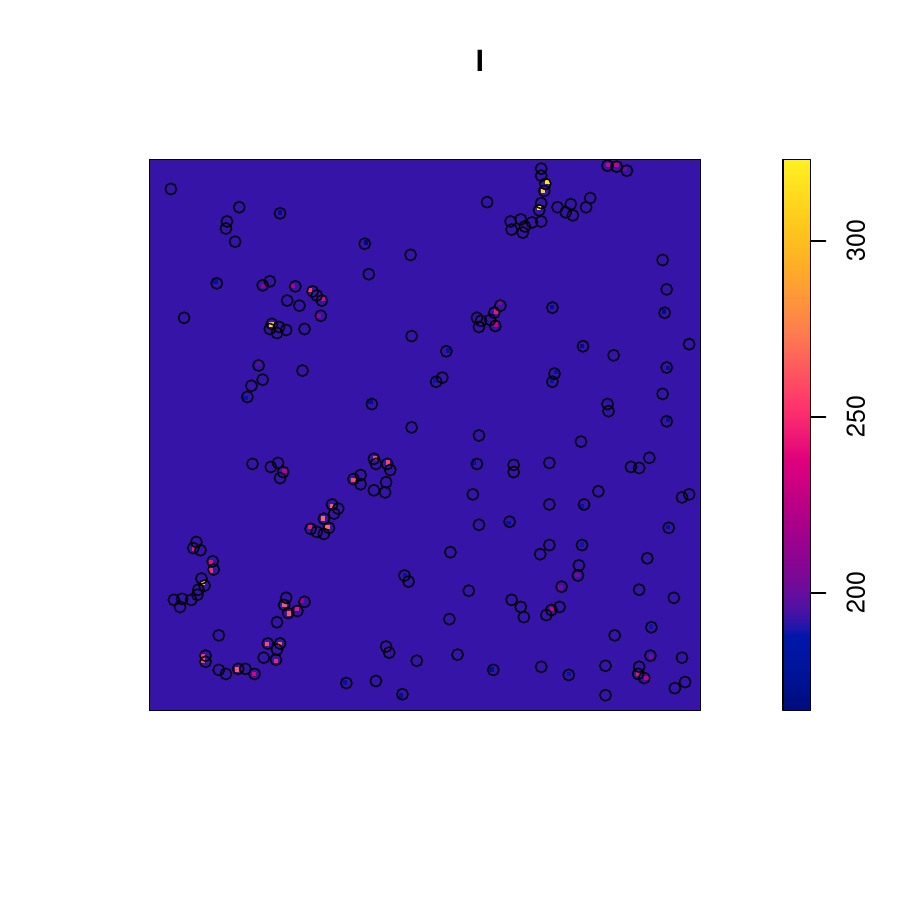}\\
        \includegraphics[width = .475\textwidth]{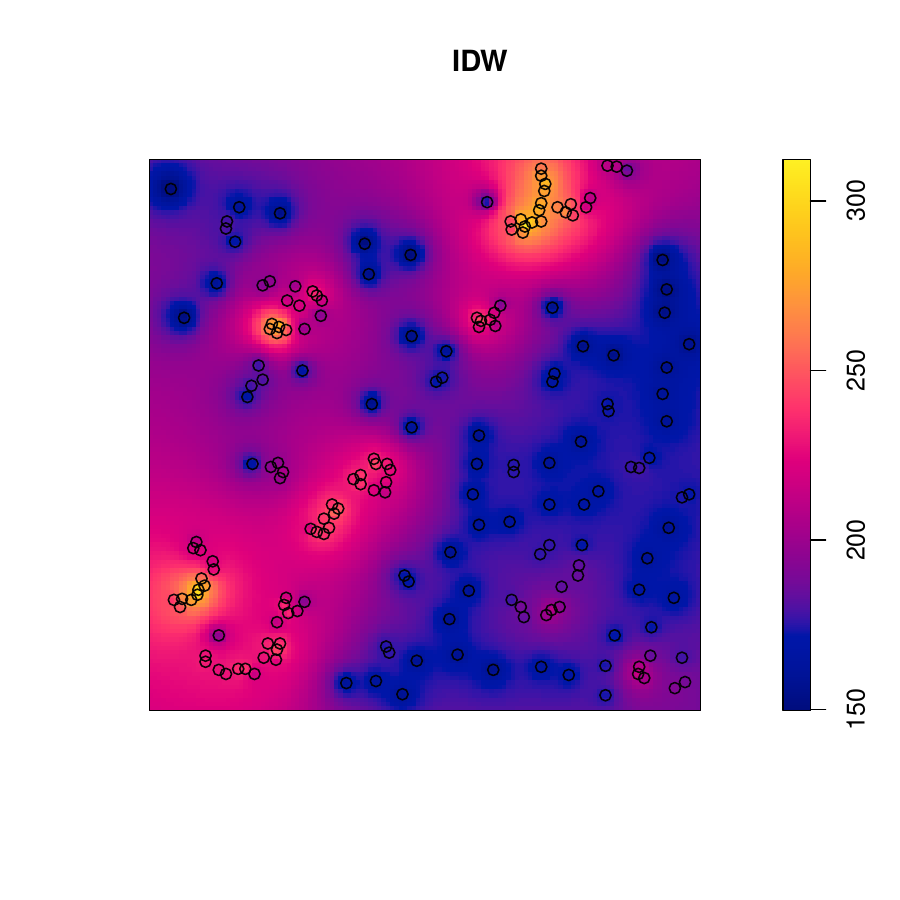}
    \includegraphics[width = .475\textwidth]{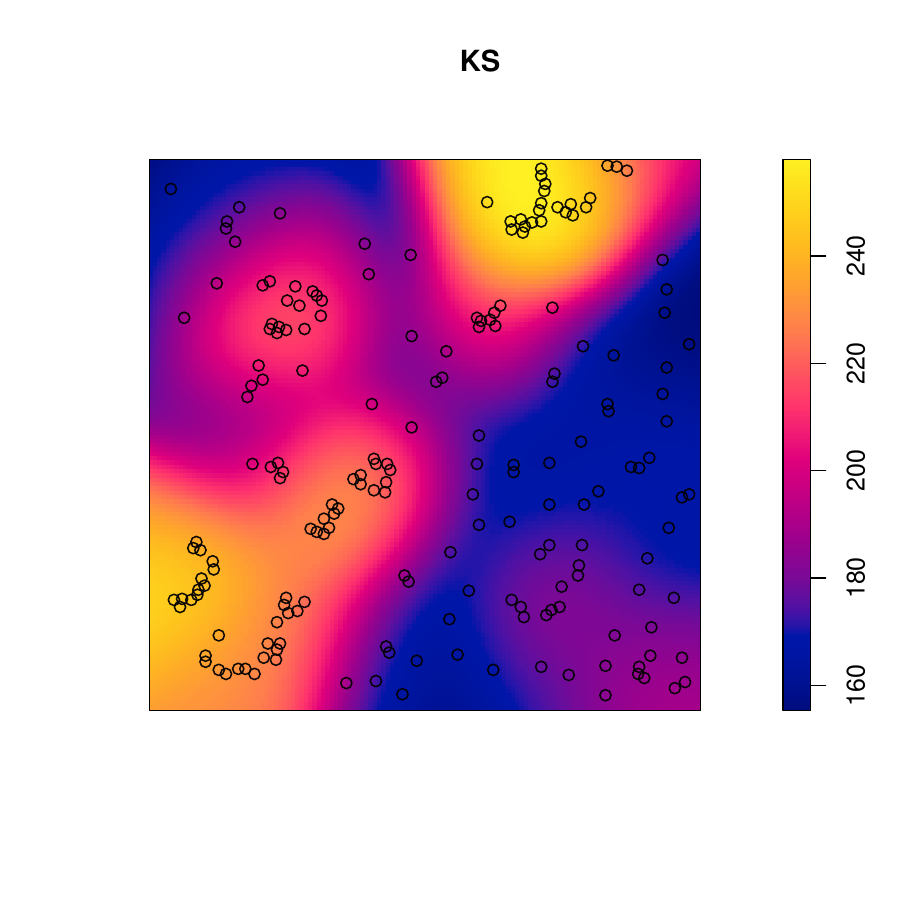}
    \caption{Predicted  intensities according to the fitted Poisson spatial models}
    \label{fig:4}
\end{figure}

Furthermore, the intensities resulting from the fitting of the four Poisson models are shown in Figure \ref{fig:4}.

It is possible to note how all the penalised models provide a non-constant fitted intensity function. In particular, moving from the \textit{indicator} method to the IDW and kernel ones, the smoothing strength increases.

\begin{figure}[H]
    \centering
    \includegraphics[width = .475\textwidth]{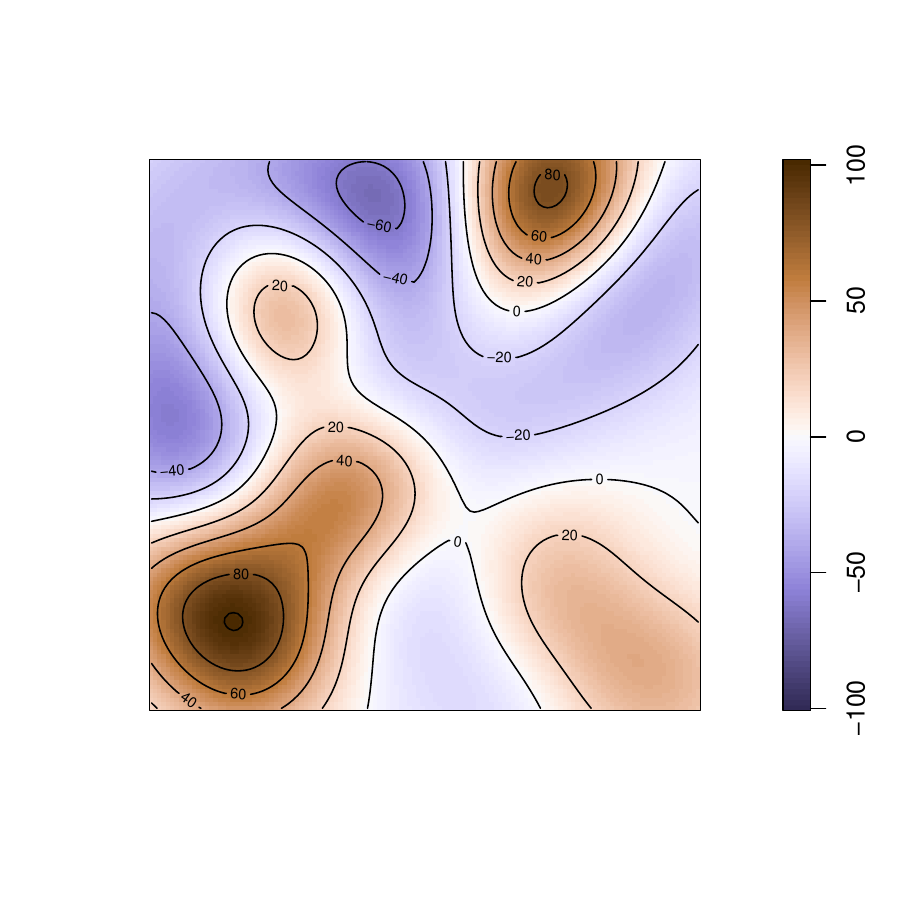}
    \includegraphics[width = .475\textwidth]{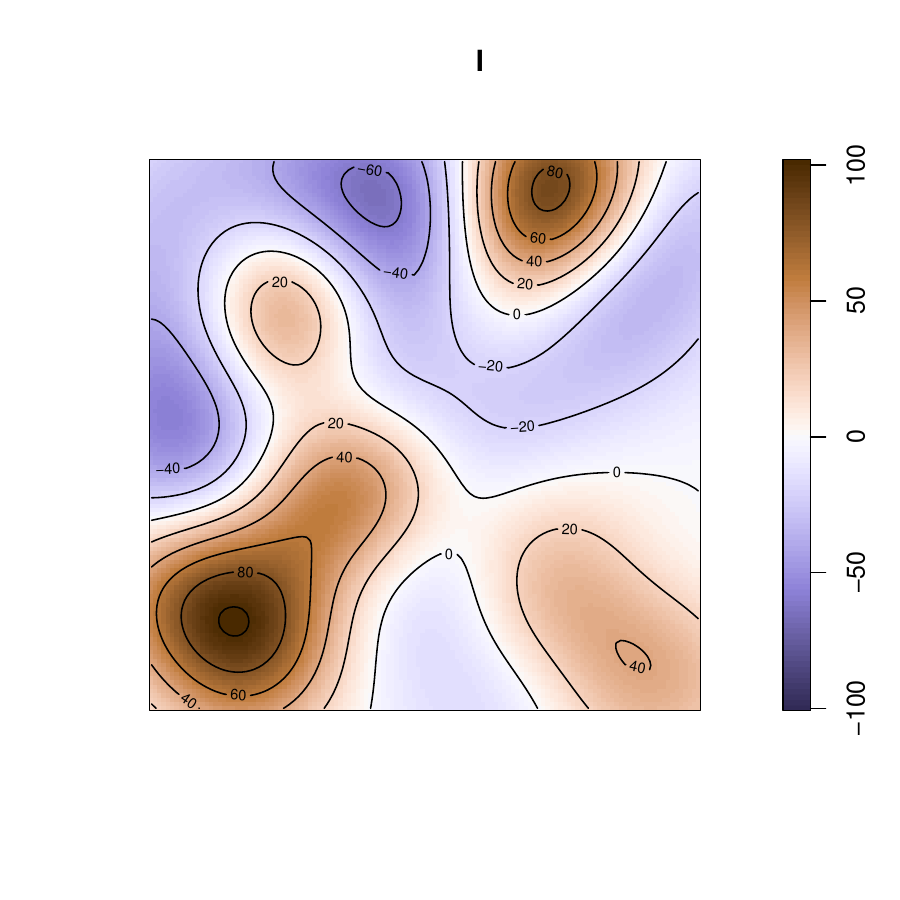}\\
        \includegraphics[width = .475\textwidth]{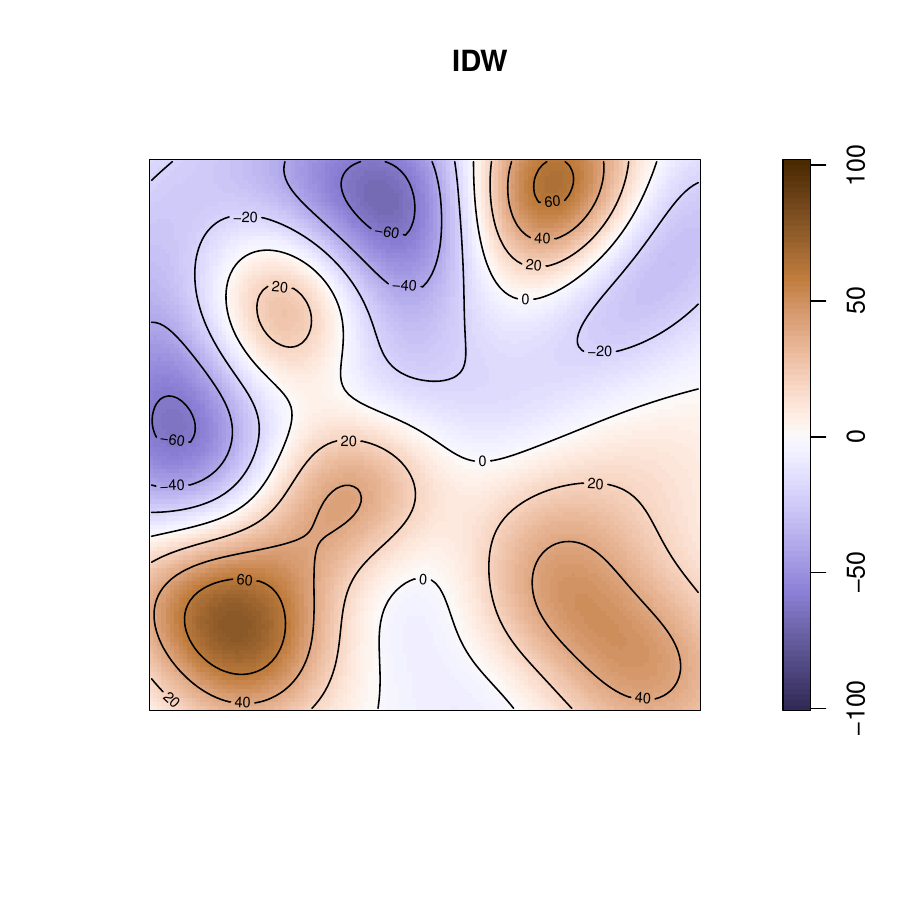}
    \includegraphics[width = .475\textwidth]{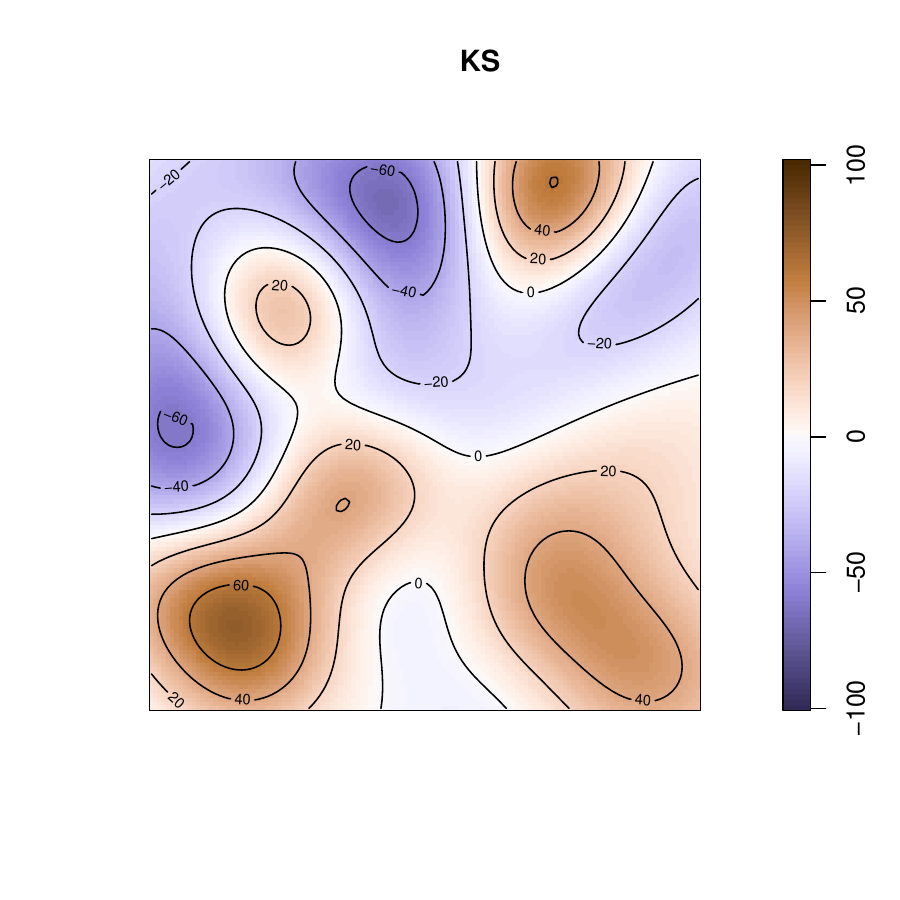}
    \caption{ Smoothed raw residuals of the fitted Poisson models}
    \label{fig:5}
\end{figure}

Raw residuals are illustrated in Figure  \ref{fig:5}. This comparison contrasts the most adaptive kernel depicted in Figure \ref{fig:22} with those fitted in Figure \ref{fig:4}.

We outline that the kernel and IDW methods obtain the best fit, and emphasise the importance of this discovery, namely the possibility of obtaining smoothing typical of non-parametric methods while retaining all the advantages of parametric fitting. These include, for example, the possibility of obtaining an information criterion and the significance of the fitted parameters.
Note that this is achieved without making any assumption about the degree of clustering of the process.

\section{Conclusions}
\label{sec:conc}


We have presented a methodological framework that exploits local second-order statistics to improve inference for complex intensity functions in point processes. By incorporating spatial dependencies through the Papangelou conditional intensity function for general Gibbs processes, we have shown the effectiveness of our approach in capturing inhomogeneity and clustering features in observed point patterns.

Our simulation results and application to real data have presented a promising performance in assessing the goodness-of-fit of the proposed method, highlighting its potential for practical applications where traditional Poisson likelihood approaches may fall short due to limited data or complex interactions. By considering local second-order characteristics, we can gain deeper insights into the local structure of point patterns, which is crucial for understanding underlying processes and making more accurate statistical inferences.

Overall, our work contributes to advancing statistical techniques for analysing spatial point patterns, and opens avenues for more robust and interpretable inference in various domains.

Future research directions may involve extending our methodology to handle additional complexities, and further exploring the theoretical properties of LISA functions in the context of point process modelling. 

An additional challenge would include the extension in the spatio-temporal domain.
Indeed, while the use of local spatio-temporal second-order summary statistics is becoming a well-established practice to describe interaction structures between points in a spatial point pattern, the use of local spatio-temporal tools was firstly advocated by \cite{siino2018testing}. They introduce Local Indicators of Spatio-temporal Association (LISTA) functions as an extension of the purely spatial LISA functions. Successively, \cite{adelfio2020some} introduced local versions of both the homogeneous and inhomogeneous spatio-temporal $K$-functions on the Euclidean space, and used them as diagnostic tools, while also retaining local information.
This spatio-temporal extension could be implemented thanks to the cubature scheme \citep{d2024preprint} recently developed to fit Poisson point processes in three dimensions. This has been implemented for general use in the \texttt{R} package \texttt{stopp} \cite{stopp}, available from the Comprehensive R Archive Network (CRAN), in order to provide a comprehensive modelling framework for spatio-temporal Poisson point processes.

\section*{Funding}
The research work of Nicoletta D'Angelo and Giada Adelfio was supported by:
 \begin{itemize}
     \item Targeted Research Funds 2024 (FFR 2024) of the University of Palermo (Italy);
     \item Mobilità e Formazione Internazionali - Miur INT project ``Sviluppo di metodologie per processi di punto spazio-temporali marcati funzionali per la previsione probabilistica dei terremoti";
     \item European Union -  NextGenerationEU, in the framework of the GRINS -Growing Resilient, INclusive and Sustainable project \\(GRINS PE00000018 – CUP  C93C22005270001).
\end{itemize}
Ottmar Cronie was supported by the Swedish research council (2023-03320) and Jorge Mateu was supported by the Ministry of Science and Innovatin (PID2022-141555OB-I00), and Generalitat Valenciana (CIAICO/2022/191).

The views and opinions expressed are solely those of the authors and do not necessarily reflect those of the European Union, nor can the European Union be held responsible for them.

\bibliography{BBB}

\end{document}